\documentclass[pdflatex,sn-mathphys-num]{sn-jnl}

\usepackage{graphicx}\usepackage{multirow}\usepackage{amsmath,amssymb,amsfonts}\usepackage{amsthm}\usepackage{mathrsfs}\usepackage[title]{appendix}\usepackage{xcolor}\usepackage{textcomp}\usepackage{manyfoot}\usepackage{booktabs}\usepackage{algorithm}\usepackage{algorithmicx}\usepackage{algpseudocode}\usepackage{listings}
\usepackage{hyperref}
\usepackage{url}

\theoremstyle{thmstyleone}
\theoremstyle{thmstyletwo}
\theoremstyle{thmstylethree}
\begin{document}

\title[SynLeaF: A Dual-Stage Multimodal Fusion Framework for Synthetic Lethality Prediction Across Pan- and Single-Cancer Contexts]{SynLeaF: A Dual-Stage Multimodal Fusion Framework for Synthetic Lethality Prediction Across Pan- and Single-Cancer Contexts}

\author[1]{\fnm{Zheming} \sur{Xing}}

\author[1]{\fnm{Siyuan} \sur{Zhou}}

\author[1]{\fnm{Ruinan} \sur{Wang}}

\author[1]{\fnm{Rui} \sur{Han}}

\author[1]{\fnm{Shiming} \sur{Zhang}}

\author[1]{\fnm{Shiqu} \sur{Chen}}

\author[1]{\fnm{Yurui} \sur{Huang}}

\author[3]{\fnm{Jiahao} \sur{Ma}}

\author[4]{\fnm{Yifan} \sur{Chen}}

\author[1]{\fnm{Xuan} \sur{Wang}}

\author[2]{\fnm{Yadong} \sur{Wang}}

\author*[1,2]{\fnm{Junyi} \sur{Li}}\email{lijunyi@hit.edu.cn}

\affil*[1]{\orgdiv{School of Computer Science and Technology}, \orgname{Harbin Institute of Technology (Shenzhen)}, \orgaddress{ \city{Shenzhen},  \state{Guang Dong} \postcode{518055}, \country{China}}}

\affil[2]{\orgdiv{Key Laboratory of Biological Bigdata, Ministry of Education}, \orgname{Harbin Institute of Technology}, \orgaddress{ \city{Harbin}, \state{Heilongjiang} \postcode{150001}, \country{China}}}

\affil[3]{\orgdiv{School of Biomedical Sciences}, \orgname{The University of Hong Kong}, \orgaddress{\state{Hong Kong SAR}, \country{China}}}

\affil[4]{\orgdiv{Departments of Mathematics and Computer Science}, \orgname{Hong Kong Baptist University}, \orgaddress{\state{Hong Kong SAR}, \country{China}}}

\abstract{
Accurate prediction of synthetic lethality (SL) is important for guiding the development of cancer drugs and therapies. SL prediction faces significant challenges in effectively fusing heterogeneous multi-source data.
Existing multimodal methods often suffer from ``modality laziness'' due to disparate convergence speeds, 
which hinders the exploitation of complementary information and causes most SL prediction models to perform poorly for both pan-cancer and single-cancer SL pair predictions. In this study, we propose SynLeaF, a dual-stage multimodal fusion framework for SL prediction in pan-cancer and single-cancer contexts. 
The framework employs a VAE-based cross-encoder with a Product of Experts mechanism to fuse four omics data types 
(gene expression, mutation, methylation, and CNV), 
simultaneously utilizing a relational graph convolutional network to capture structured gene representations from biomedical knowledge graphs. 
To mitigate modality laziness, SynLeaF introduces a dual-stage training mechanism that employs a feature-level knowledge distillation. 
In extensive experiments across eight specific cancer types and a pan-cancer dataset, 
SynLeaF achieved superior performance in 17 of the 19 scenarios.
Ablation studies and gradient analyses further validate the critical contributions 
of the proposed fusion and distillation mechanisms for model robustness and generalization.
To facilitate community use, a web server is available at \url{https://synleaf.bioinformatics-lilab.cn}.
}


\maketitle

\section{Introduction}\label{sec1}

Synthetic Lethality (SL) characterizes a specific genetic relationship 
wherein the deficiency of an individual gene remains viable for the cell, 
whereas concurrent impairment or inactivation of a gene pair results in cell death \cite{Huang2020}.
As a promising targeted anti-cancer therapy, 
SL can eliminate malignant cells while preserving healthy tissues \cite{Ashworth2018}, 
and expands druggable targets for genes that are difficult to target directly \cite{Wang2022}.
A classic clinical success is the PARP inhibitors, approved by the FDA in 2014 
for ovarian cancer with BRCA1/2 mutations \cite{Topatana2020, Lord2017}. 
Recently, computationally designed aptamers were shown to induce SL 
by blocking the RAD51--BRCA2 interaction \cite{Milordini2025}. 
ADAR1 was also identified as a new SL target in BRCA-mutant tumors, 
where its inhibition activates innate immunity via autocrine interferon poisoning \cite{Chabanon2025}.
These advances continue to expand druggable targets and accelerate their translation into therapies \cite{Goncalves2026}.

Despite this potential, identifying clinically relevant SL pairs remains a challenge ~\cite{ONeil2017}.
Wet-lab screens, including yeast, RNAi, and CRISPR, 
are accurate but costly, time-consuming, and suffer from significant off-target effects \cite{Hao2021, KaramiFath2025}.
The sheer volume of possible gene pairs makes exhaustive screening infeasible \cite{Horlbeck2018}.
To overcome the limitations of wet-lab methods, computational methods, 
which can be grouped into statistics-based, network-based, traditional machine learning, and deep learning methods, 
have emerged as effective complements \cite{Wang2022}.

Statistics and network approaches for synthetic lethal prediction rely on hypotheses and domain knowledge 
and often ignore other data types such as sequences or functional attributes \cite{Nijman2011}.
Traditional machine learning methods integrate multi-source data 
but still depend on feature engineering, which can introduce noise \cite{Das2019, Wan2020}.
To address these limitations, deep learning methods have been introduced to automatically learn complex representations.
Deep learning methods learn complex representations automatically.
Early network representation models such as GRSMF and SL2MF use matrix factorization \cite{Huang2019, Liu2020}.
GRSMF introduces graph-regularized self-representation to alleviate data sparsity, 
and SL2MF uses logistic matrix factorization with different weights for known and unknown pairs.
However, matrix factorization methods are essentially a form of shallow embedding 
and may not fully exploit the structural information and node features of the network \cite{Hamilton2020}.

The application of Graph Neural Networks (GNN) has significantly enhanced the accuracy of predicting synthetic lethality. 
DDGCN \cite{Cai2020} uses a dual-dropout strategy to address sparsity and overfitting, 
and GCATSL \cite{Long2021} applies dual attention at the node and feature levels.
However, these GNN-based methods mainly rely on homogeneous networks, 
which only contain gene nodes and possess limited expressive power.
The incorporation of Knowledge Graphs (KG) has further broadened 
the use of deep learning techniques for synthetic lethality prediction.
Basically, a KG represents a heterogeneous network that contains multiple types of entities 
(such as genes, pathways, diseases, etc.) and relations, 
that can provide richer biological background information \cite{Ye2021}.
KG4SL \cite{Wang2021} automatically generates gene features using a knowledge graph convolutional network.
PiLSL \cite{Liu2022} adopts a gene pair-based method, 
extracting enclosing subgraphs from the knowledge graph 
to capture pairwise interactions between gene pairs.
SLGNN \cite{Zhu2023} models relation combinations as factors and improves the interpretability 
of the model through node- and factor-level attention mechanisms.
KR4SL \cite{Zhang2023} introduces a reasoning method based on paths, 
utilizing relational digraphs to extract structural semantic information within the knowledge graph. 
Meanwhile, MPASL \cite{Zhang2024} proposed a hybrid interaction framework 
involving gene-entity and entity-entity relationships 
to improve gene representations from various viewpoints.

Although knowledge graphs provide rich structured biological knowledge, 
KGs solely from a single modality often cannot fully capture the complex mechanisms of synthetic lethality. 
Consequently, fusing data from multiple sources has emerged as a crucial approach 
to enhance the accuracy of synthetic lethality prediction.
PTGNN \cite{Long2022} uses Convolutional Neural Network (CNN) features from protein sequences and a graph reconstruction pre-training task 
on Protein-Protein Interaction and Gene Ontology (GO) graphs. 
PiLSL \cite{Liu2022} integrates explicit omics features with KG embeddings. 
TARSL \cite{Li2025} applies non-negative matrix tri-factorization with triple attention, 
and Struct2SL \cite{Huang2025} combines protein sequences, protein 3D structures, and PPI networks.
Although these methods improve prediction accuracy by introducing multi-source information, 
they face a common and critical limitation, as in pure KG methods, the neglect of \emph{context-specificity}.

The realization of generalized synthetic lethal effects 
is often impeded by specific tumor-associated factors, 
including cellular heterogeneity, metabolic status, and the complexities of the tumor microenvironment \cite{Previtali2024}.
Many synthetic lethal interactions are observed in only a few specific cancers.
An investigation utilizing CRISPR-Cas9 screening to pinpoint synthetic lethal gene pairs across three cell lines revealed a minimal intersection: 
merely 10\% of the interactions were shared between any two lines, while no single pair was consistently observed across all three \cite{Shen2017, Tang2022}.
Adequately addressing tumor heterogeneity can mitigate certain challenges 
associated with the translation of synthetic lethality strategies 
from cellular models to \textit{in vivo} systems and, ultimately, clinical applications \cite{Ryan2018}.
Context-specific synthetic lethal effects have garnered significant interest within the medicinal chemistry community, 
as investigating cancer-specific synthetic lethality offers novel avenues for pharmaceutical development.
Although several studies have attempted to address the specificity problem, 
they still exhibit significant shortcomings.
While ELISL \cite{Tepeli2023} pioneered the integration of context-free protein sequence 
associations with context-specific omics data, 
its reliance on shallow models, such as random forests, limits its ability to capture the non-linear relationships in high-dimensional data.
Based on cancer-specific positive synthetic lethality datasets, 
SLGNNCT \cite{Chen2024} divides the knowledge graph into different subgraphs 
and performs cancer-specific synthetic lethality prediction on the small knowledge graphs separately via factor attention modeling, such as SLGNN.
However, the resulting knowledge subgraphs have very few nodes, 
and the small graph size leads to poor generalization performance in deep learning models.

To address these obstacles, we propose SynLeaF, 
a dual-stage multimodal fusion framework for SL prediction across pan- and single-cancer contexts.
The method combines a Cross-VAE with Product-of-Experts early fusion for four omics modalities and an RGCN-based KG encoder, and introduces an adaptive two-stage fusion/training paradigm (Uni-Modal Teacher, UMT; Uni-Modal Ensemble, UME) to mitigate modality laziness~\cite{NEURIPS2022_662b1774, Kutuzova2021, Schlichtkrull2017, Du2023}.
In extensive experiments covering eight specific cancer types and a pan-cancer dataset, 
SynLeaF demonstrated excellent generalization capability and robustness,
surpassing existing state-of-the-art techniques across the majority of core evaluation metrics.

\section{Results}\label{sec2}

\subsection{SynLeaF: A Dual-Stage Multimodal Fusion Framework}

\begin{figure*}[htbp]
    \centering
    \includegraphics[width=1\textwidth]{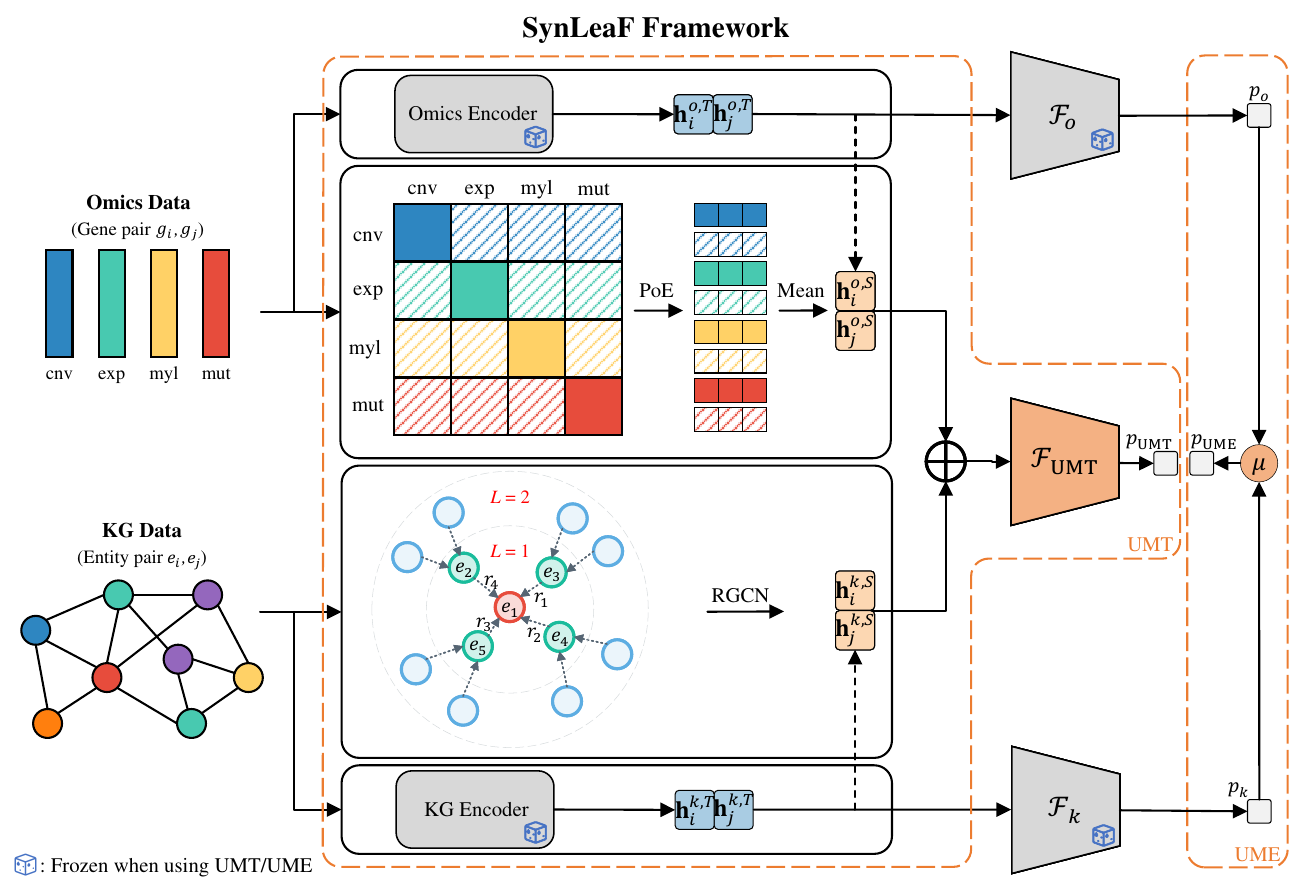}      \caption{
\textbf{Overview of the SynLeaF Framework}.
SynLeaF constitutes a dual-stage multimodal integration architecture designed for synthetic lethality prediction, 
taking cancer-specific omics profiles and biomedical knowledge graphs as inputs.
The \textbf{Omics Encoder} uses a cross-encoder architecture based on a Variational Autoencoder (VAE)
and performs early fusion on four types of omics data: copy number variation (cnv), gene expression (exp), DNA methylation (myl), and mutation (mut),
through a Product of Experts (PoE) mechanism.
The \textbf{Knowledge Graph Encoder} utilizes a Relational Graph Convolutional Network (RGCN) 
to extract structural features associated with genes within the biological network.
With these encoders, SynLeaF first independently pre-trains two unimodal models 
and then constructs two base estimators respectively under two complementary fusion strategies in further training.
For both the UMT and UME strategies, the parameters of the pre-trained unimodal encoders are strictly frozen. 
Specifically, under the \textbf{UMT} strategy, SynLeaF treats the pre-trained unimodal encoders as teachers,
guiding the training of a multimodal student model through feature-level knowledge distillation.
Under the \textbf{UME} strategy, SynLeaF directly integrates the prediction results ($p_o$ and $p_k$) from the two pre-trained unimodal models.
Here, $p_o$ and $p_k$ denote the predicted probabilities from the Omics and Knowledge Graph models, respectively.
Finally, SynLeaF adaptively selects the optimal strategy between UMT and UME according to observed validation efficacy.
     }
    \label{fig:model_architecture} \end{figure*}

As shown in Figure \ref{fig:model_architecture}, SynLeaF is a dual-stage multimodal fusion framework,
designed for the prediction of cancer synthetic lethality by integrating heterogeneous data from multiple sources.
The model accepts two streams of input: first, four types of omics data, including gene expression, mutation, DNA methylation, and copy number variation (CNV) \footnote{
    Due to data source limitations, 
    the pan-cancer dataset uses only three types of omics 
    and does not include DNA methylation data.
};
and second, a biomedical KG that contains various entities such as genes, pathways, and diseases, along with their relationships.
The omics data are encoded through a VAE-based early fusion module,
while the KG is processed by an RGCN to extract structured representations.

The high-dimensional nature and significant modal heterogeneity of omics data hinder the effective fusion of multi-source information.
Drawing inspiration from the framework established by \citet{Wang2024}, we employed a Variational Autoencoder (VAE) architecture \cite{NEURIPS2022_662b1774}
and innovatively designed a Cross-VAE-based early fusion module.
As shown in the Omics Encoder module of Figure \ref{fig:model_architecture}, unlike traditional feature concatenation, our method constructs an $N \times N$ encoder matrix,
which skillfully balances intra-modal feature self-learning (via diagonal autoencoders) with inter-modal interactive inference (via off-diagonal cross-encoders).
By introducing the Product of Experts (PoE) mechanism \cite{Kutuzova2021} to aggregate multi-source information,
SynLeaF can generate unified and robust gene representations in the latent space.
Notably, this architecture naturally addresses the problem of missing data through its cross-inference mechanism.
Even if some omics data are unavailable, the model can still reconstruct the missing features based on other modalities.
This greatly improves the model's generalization ability when dealing with fragmented clinical data.

In multimodal learning, features are generally categorized into two distinct classes. 
The former comprises unimodal attributes acquirable via isolated training, 
whereas the latter consists of paired characteristics 
that necessitate cross-modal interaction for extraction.
Optimally, the objective is for a multimodal model to capture paired features via cross-modal mechanisms, 
while ensuring that it also learns sufficient unimodal features.
However, \citet{Du2023} found that traditional multimodal late-fusion training methods suffer from the problem of ``Modality Laziness''.
This means that encoders trained jointly on multiple modalities perform worse on unimodal feature learning than encoders trained unimodally, 
and this phenomenon is particularly evident in tasks where unimodal priors are meaningful.
To mitigate this problem, \citet{Du2023} proposed two complementary multimodal fusion strategies, designated as Uni-Modal Teacher (UMT) and Uni-Modal Ensemble (UME),
and achieved good performance in multimodal audio-visual classification tasks.
The UMT strategy uses pre-trained unimodal encoders as ``teachers'' to guide a multimodal ``student'' model
to learn the teachers' feature representations through feature-level Knowledge Distillation.
The UME strategy, on the other hand, directly integrates the prediction results of the two pre-trained unimodal models.

Inspired by the UMT/UME framework, this paper proposes a dual-stage training strategy, 
which is adapted and extended for the characteristics of the omics-graph bimodal setting 
to ensure the slow modality is fully trained.
During the initial stage, we conduct independent pre-training for both the omics encoder and the knowledge graph encoder
to ensure that each unimodal encoder can fully learn the feature representations within its modality.
During the second stage, considering the significant differences in inter-modal interactions 
across different cancer types and data splitting strategies, 
we designed an adaptive selection mechanism.
After training, the model automatically selects the best strategy between UMT and UME 
based on the Area Under the ROC Curve (AUC) metric obtained from the validation dataset.
UMT usually performs better 
when the unimodal features from both omics and the knowledge graph are strong,
and the cross-modal interaction provides additional information.
However, when one modality is clearly dominant or the cross-modal interaction introduces noise, 
the simple ensemble strategy of UME is more effective.

\subsection{SynLeaF Surpasses Current State-of-the-Art Techniques in Both Pan-Cancer and Cancer-Specific Settings}

To evaluate the effectiveness and robustness of SynLeaF, 
we conducted a comprehensive comparison with four state-of-the-art methods on a collection of datasets, 
which includes eight cancer-specific datasets and one large pan-cancer dataset, under various splitting settings.
The eight specific cancer types are: 
breast cancer (BRCA), cervical cancer (CESC), 
colon cancer (COAD), kidney renal clear cell carcinoma (KIRC), 
acute myeloid leukemia (LAML), lung adenocarcinoma (LUAD), 
ovarian cancer (OV), and skin cutaneous melanoma (SKCM).
These settings include CV1 (Random Split), CV2 (Semi-New Gene Split), and CV3 (New Gene Split) \cite{Liu2022}.
The four state-of-the-art methods compared were SLGNN, ELISL, PTGNN, and MPASL.

To guarantee a fair evaluation, we standardized the data loading part for all baseline methods to make sure they all used the same datasets, 
including the SL gene pairs, protein sequences, omics features, and knowledge graph data.
The implementation details for baselines and all training configurations 
are described in the Supplementary Material (see \textit{Experimental Design}).
Empirical outcomes indicate that SynLeaF exhibits a distinct performance edge.
On the core comparison metrics across a total of 19 scenarios, SynLeaF achieved state-of-the-art (SOTA) results in 17 of these instances.

\begin{table*}[!htbp]
\caption{Experiment results on cancer-specific datasets. 
The standard deviations of the readings are reported in parentheses.
The best performing method is highlighted in \textbf{bold} and the second best is \underline{underlined}.
The last column indicates the improvement made by SynLeaF.}
\label{tab:comp_exp_single}
\tiny
\tabcolsep=0pt
\begin{tabular*}{\textwidth}{@{\extracolsep{\fill}}lcccccccr@{\extracolsep{\fill}}}
\toprule
Cancer & Split & Metric & SLGNN & ELISL & PTGNN & MPASL &  SynLeaF & $\uparrow$(\%) \\
\midrule

BRCA&CV1&AUC&0.8847(0.0130)&0.9041(0.0104)&\underline{0.9453(0.0093)}&0.8187(0.0173)&\textbf{0.9654(0.0038)}&2.13\\
&&AUPR&0.8949(0.0169)&0.9196(0.0079)&\underline{0.9600(0.0084)}&0.8508(0.0118)&\textbf{0.9743(0.0020)}&1.49\\
&CV2&AUC&0.7147(0.1036)&0.7615(0.0470)&\underline{0.8153(0.0911)}&0.6660(0.0701)&\textbf{0.8474(0.0502)}&3.94\\
&&AUPR&0.7358(0.1114)&0.7894(0.0664)&\underline{0.8316(0.0973)}&0.6888(0.0763)&\textbf{0.8707(0.0563)}&4.70\\
CESC&CV1&AUC&0.6251(0.0795)&0.7136(0.0797)&\underline{0.7765(0.0480)}&0.7321(0.0946)&\textbf{0.8136(0.0505)}&4.78\\
&&AUPR&0.6543(0.0846)&0.7466(0.0484)&\underline{0.7915(0.0621)}&0.7299(0.0847)&\textbf{0.8180(0.0650)}&3.35\\
&CV2&AUC&0.5957(0.0489)&0.5280(0.0374)&\underline{0.6834(0.0948)}&0.6297(0.0750)&\textbf{0.6845(0.0444)}&0.16\\
&&AUPR&0.5834(0.0443)&0.5753(0.0384)&\underline{0.6946(0.0967)}&0.6412(0.0272)&\textbf{0.7190(0.0530)}&3.51\\
COAD&CV1&AUC&0.5933(0.0291)&\underline{0.6894(0.0150)}&0.6212(0.0498)&0.6278(0.0172)&\textbf{0.7162(0.0263)}&3.89\\
&&AUPR&0.5979(0.0416)&\underline{0.6645(0.0382)}&0.6071(0.0592)&0.6535(0.0257)&\textbf{0.7088(0.0287)}&6.67\\
&CV2&AUC&0.5404(0.0237)&\underline{0.6206(0.0391)}&0.5157(0.0364)&0.5702(0.0385)&\textbf{0.6317(0.0244)}&1.79\\
&&AUPR&0.5397(0.0280)&\underline{0.6064(0.0316)}&0.5181(0.0279)&0.6031(0.0397)&\textbf{0.6120(0.0316)}&0.92\\
KIRC&CV1&AUC&0.6459(0.1381)&0.6754(0.0886)&\textbf{0.7161(0.0957)}&0.6739(0.0601)&\underline{0.6940(0.0990)}&-\\
&&AUPR&0.6668(0.1067)&\underline{0.7238(0.0806)}&\textbf{0.7327(0.0908)}&0.7059(0.0835)&0.7162(0.0855)&-\\
&CV2&AUC&0.5417(0.1334)&\textbf{0.6599(0.1128)}&\underline{0.6430(0.1319)}&0.6317(0.1601)&0.5822(0.0573)&-\\
&&AUPR&0.5647(0.1054)&\textbf{0.6646(0.1106)}&\underline{0.6441(0.1123)}&0.6191(0.1975)&0.5978(0.0770)&-\\
LAML&CV1&AUC&0.5793(0.0224)&0.6267(0.0061)&\underline{0.6960(0.0303)}&0.6306(0.0256)&\textbf{0.6980(0.0156)}&0.29\\
&&AUPR&0.5914(0.0247)&0.6310(0.0177)&\underline{0.6925(0.0533)}&0.6605(0.0151)&\textbf{0.7002(0.0239)}&1.11\\
&CV2&AUC&0.5467(0.0258)&0.5810(0.0218)&\underline{0.6188(0.0366)}&0.5977(0.0468)&\textbf{0.6296(0.0195)}&1.75\\
&&AUPR&0.5648(0.0200)&0.5930(0.0222)&0.6220(0.0518)&\underline{0.6297(0.0326)}&\textbf{0.6378(0.0205)}&1.29\\
LUAD&CV1&AUC&0.8254(0.0229)&0.8513(0.0295)&\underline{0.8865(0.0254)}&0.7945(0.0584)&\textbf{0.9000(0.0259)}&1.52\\
&&AUPR&0.8372(0.0132)&0.8571(0.0336)&\underline{0.8753(0.0356)}&0.7951(0.0471)&\textbf{0.8873(0.0273)}&1.37\\
&CV2&AUC&0.5858(0.1520)&0.7465(0.0920)&\underline{0.7678(0.0982)}&0.6336(0.1427)&\textbf{0.8161(0.0854)}&6.29\\
&&AUPR&0.6365(0.1477)&0.7407(0.1208)&\underline{0.7676(0.1140)}&0.6636(0.1326)&\textbf{0.7924(0.1110)}&3.23\\
OV&CV1&AUC&0.9201(0.0209)&0.7790(0.0683)&\underline{0.9824(0.0116)}&0.8555(0.0341)&\textbf{0.9827(0.0144)}&0.03\\
&&AUPR&0.9426(0.0166)&0.7812(0.0867)&\underline{0.9590(0.0384)}&0.8325(0.0413)&\textbf{0.9855(0.0103)}&2.76\\
&CV2&AUC&0.5894(0.0828)&0.7087(0.0300)&\underline{0.7416(0.0813)}&0.6846(0.0601)&\textbf{0.7990(0.0850)}&7.74\\
&&AUPR&0.6136(0.0598)&0.6812(0.0202)&0.6694(0.0854)&\underline{0.6867(0.0620)}&\textbf{0.7252(0.1149)}&5.61\\
SKCM&CV1&AUC&0.6692(0.1494)&\underline{0.6871(0.0248)}&0.6469(0.0788)&0.5849(0.1209)&\textbf{0.8088(0.0307)}&17.71\\
&&AUPR&0.7106(0.1710)&\underline{0.7173(0.0451)}&0.6815(0.0737)&0.6745(0.0609)&\textbf{0.8471(0.0436)}&18.10\\
&CV2&AUC&0.6777(0.0862)&0.6302(0.1424)&\underline{0.7138(0.2257)}&0.5760(0.1715)&\textbf{0.7630(0.1576)}&6.89\\
&&AUPR&0.7059(0.0661)&0.6216(0.1371)&\underline{0.7251(0.2111)}&0.6150(0.1551)&\textbf{0.7876(0.1566)}&8.62\\
\bottomrule
\end{tabular*}
\end{table*}

\begin{table*}[!htbp]
\caption{Comparison of experimental results on the pan-cancer dataset}
\label{tab:comp_exp_pan}
\tiny
\tabcolsep=0pt
\begin{tabular*}{\textwidth}{@{\extracolsep{\fill}}lccccccr@{\extracolsep{\fill}}}
\toprule
Cancer & Split & Metric & SLGNN &  PTGNN & MPASL &  SynLeaF & $\uparrow$(\%) \\
\midrule

pan&CV1&AUC&\underline{0.9550(0.0021)}&0.9315(0.0011)&0.9336(0.0037)&\textbf{0.9652(0.0012)}&1.07\\
&&AUPR&\underline{0.9616(0.0018)}&0.9386(0.0024)&0.9425(0.0033)&\textbf{0.9669(0.0008)}&0.55\\
&&F1&\underline{0.8964(0.0023)}&0.8894(0.0014)&0.8692(0.0041)&\textbf{0.9099(0.0018)}&1.51\\

&CV2&AUC&\underline{0.7736(0.0305)}&0.7684(0.0657)&0.4900(0.0234)&\textbf{0.8624(0.0236)}&11.48\\
&&AUPR&0.8050(0.0255)&\underline{0.8104(0.0448)}&0.6118(0.0232)&\textbf{0.8754(0.0226)}&8.02\\
&&F1&0.6006(0.0394)&\underline{0.7189(0.0433)}&0.3858(0.1325)&\textbf{0.7955(0.0246)}&10.66\\

&CV3&AUC&\underline{0.5757(0.0367)}&0.5128(0.0576)&0.5379(0.0469)&\textbf{0.7407(0.0271)}&28.66\\
&&AUPR&\underline{0.6050(0.0414)}&0.5213(0.0494)&0.5416(0.0413)&\textbf{0.7611(0.0417)}&25.80\\
&&F1&0.1163(0.0944)&\underline{0.6673(0.0012)}&0.0000(0.0000)&\textbf{0.7153(0.0156)}&7.19\\

\bottomrule
\end{tabular*}
\end{table*}

Table \ref{tab:comp_exp_single} details the comparative outcomes across cancer-specific datasets,
where we observe that SynLeaF has a significant lead in most cancer types.
Notably, regarding the SKCM dataset, when evaluated against the second-best model, 
SynLeaF achieved huge improvements of 17.71\% and 6.89\% under the CV1 and CV2 settings, respectively.
Since the ELISL method relies on cancer-specific clinical omics data and cell line omics data, 
it cannot be applied to the pan-cancer synthetic lethality prediction task.
Therefore, it was not included in the comparison.
The experimental data confirm that our proposed method achieved the highest performance metrics under all splitting strategies on the pan-cancer dataset.
Particularly in the CV3 (zero-shot) setting, which simulates the prediction of unknown genes, SynLeaF still achieved an AUC of 0.7407,
representing an improvement of up to 28.66\% over the second-place SLGNN.
PTGNN showed strong competitiveness in the single-cancer experiments, 
ranking second after SynLeaF on most metrics.
However, this advantage has its limitations.
By looking at Table \ref{tab:comp_exp_single} and Table \ref{tab:comp_exp_pan} together, we can see that
although PTGNN can fit the single distribution of a specific cancer, 
its performance dropped significantly when faced with the pan-cancer scenario, 
which has more complex data and greater distribution differences.
In contrast, SynLeaF demonstrated an all-around adaptability,
as it maintained the best performance across both the smaller single-cancer datasets and the larger pan-cancer dataset.

Although SynLeaF performed excellently on the vast majority of datasets, 
the framework failed to secure the leading position on the KIRC dataset.
We conducted an in-depth analysis of this phenomenon 
and found that the core reason is the extremely small sample size (only about 120 samples).
This caused a serious distribution shift between the validation and test sets, 
which in turn led to a misjudgment by the adaptive selection strategy.
This statistical deviation caused the validation-set-based adaptive selection strategy to be conservative.
Specifically, the sub-modules of SynLeaF actually have the potential to reach SOTA on KIRC, especially in the CV1 setting.
On the testing partition, the UMT branch achieved an AUC of 0.7218 ($\pm$0.0923) and an AUPR of 0.7246 ($\pm$0.0737), 
metrics that align closely with the efficacy exhibited by the top-ranked PTGNN.
This result indicates that in small-sample and highly heterogeneous cancer datasets, 
relying solely on validation set metrics for model selection can be risky.
The result on KIRC reveals that how to overcome distribution shift and evaluation bias in extremely data-sparse scenarios 
remains a common challenge for the entire field of computational biology.

\subsection{The Dual-Stage Adaptive Fusion Strategy Effectively Addresses Modality Dependency and Heterogeneity Challenges}

To investigate the contribution of multimodal data to synthetic lethality prediction 
and to validate the effectiveness of the SynLeaF dual-stage fusion strategy,
we executed an in-depth ablation analysis focusing on the Only Omics, Only KG, and the full version of SynLeaF.

\subsubsection{Multimodal Integration Shows an ``Envelope Effect'' and Robustness Superior to Unimodal Benchmarks}

\begin{figure}[!htbp]
    \centering
    \includegraphics[width=1\textwidth]{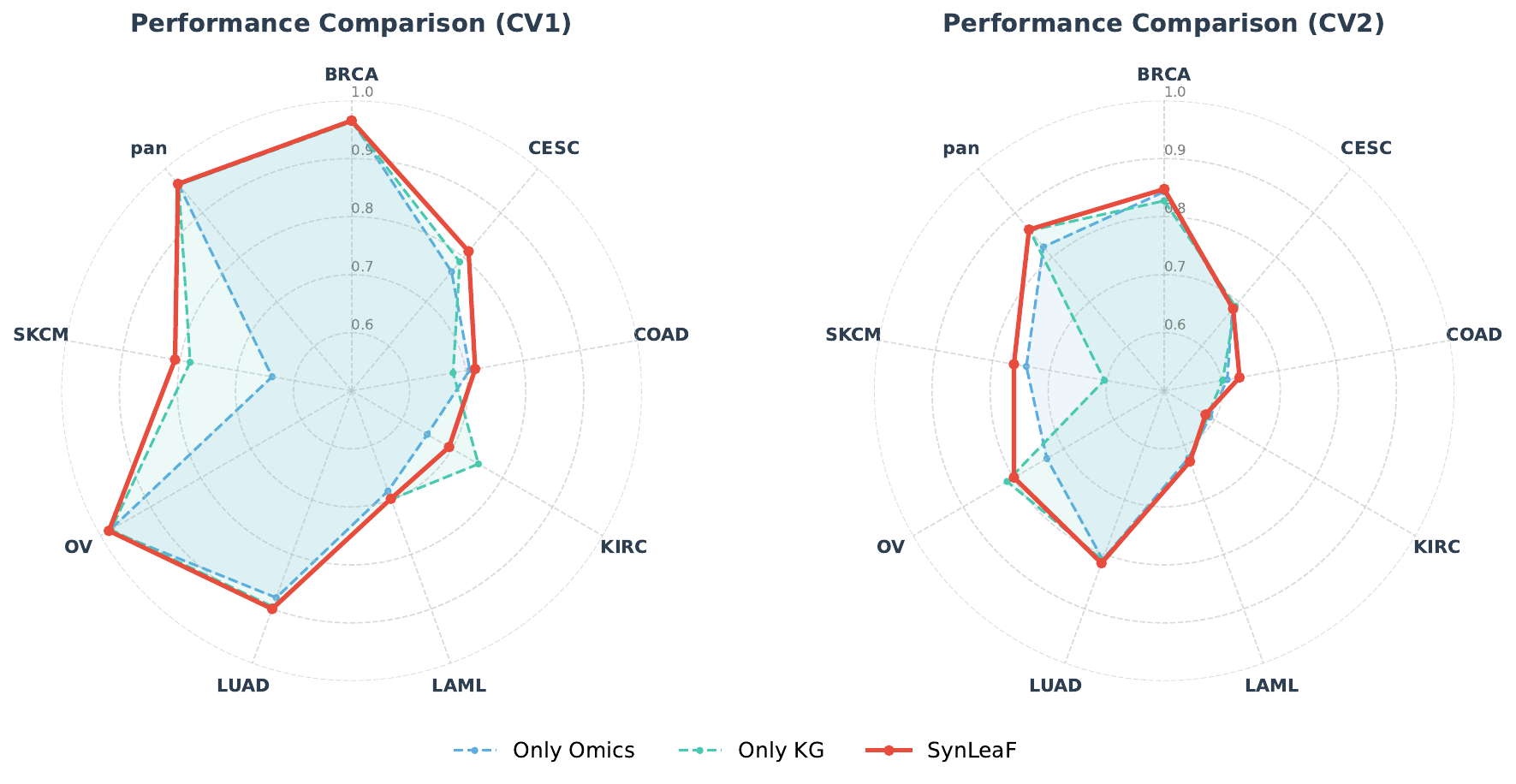}     \caption{
\textbf{Radar chart comparing the performance of SynLeaF and unimodal baseline variants.}
This chart shows the AUC performance comparison of SynLeaF against the two unimodal baseline variants, 
Only Omics and Only KG, on the pan-cancer and eight cancer-specific datasets,
under the two data splitting strategies of CV1 and CV2.
The performance curve of SynLeaF forms an ``envelope effect'' over the unimodal models on almost all datasets, 
demonstrating the consistent advantage of multimodal fusion.
}
    \label{fig:ablation_radar} \end{figure}

As shown in Figure \ref{fig:ablation_radar}, SynLeaF forms a clear ``envelope effect'' 
over the unimodal models, indicating stable gains from multimodal fusion.
Notably, modal advantages shift drastically depending on the data split. 
For SKCM in CV1, KG provides the main predictive signal (AUC $\approx$ 0.78) compared to Omics (AUC $\approx$ 0.64).
However, in CV2 (unseen genes), a modal reversal occurs: KG drops sharply to 0.60 due to sparse graph connections, 
while Omics rises to 0.74. 
Despite this fluctuation, SynLeaF adaptively shifts its focus to Omics, maintaining a robust AUC of 
0.76.

It is worth noting that this modal advantage shows a different pattern from the pan-cancer perspective.
In the CV1 setting on the pan-cancer dataset, both Omics and KG showed very high performance.
But in the more challenging zero-shot condition of the CV3 setting, the situation was reversed.
The performance of Only Omics dropped significantly to 0.6259, while Only KG still maintained a robust AUC of 0.7391.
This reveals that in complex scenarios that span multiple cancer types, 
simple omics features are easily affected by heterogeneity noise.
The global knowledge graph, on the other hand, provides a stronger inductive bias through the biological network topology, 
thus enabling structural inference on completely unknown genes.

In summary, the local case of SKCM and the global case of pan-cancer together prove that
no single modality can excel under all splits, and SynLeaF's multimodal fusion mechanism is an essential approach to handle this complexity.

\subsubsection{The Adaptive Selection Mechanism Captures the Differential Modality Dependency of Different Cancers}

Although multimodal fusion is generally effective, 
we observed a key phenomenon that different cancer datasets show very different modality dependency under different splitting strategies.
In the synthetic lethality prediction task, deep cross-modal interaction is very important in some scenarios, 
while in other scenarios, forced interaction can instead introduce noise.
Therefore, SynLeaF introduces an adaptive selection mechanism to address the significant differences in inter-modal interactions 
under different data splitting strategies.
To accommodate the heterogeneity within data distributions, 
the model identifies the most effective fusion path by evaluating outcomes on the validation set.
We plotted bar charts to evaluate the effectiveness of the modal fusion strategies. 
Due to space limitations, this section only discusses two representative datasets, CESC CV1 and COAD CV1,
as shown in Figure \ref{fig:ablation_bar}. 
Detailed outcomes of the full experiments are presented in Supplementary Figure S3.

\begin{figure}[!htbp]
    \centering
    \includegraphics[width=1\textwidth]{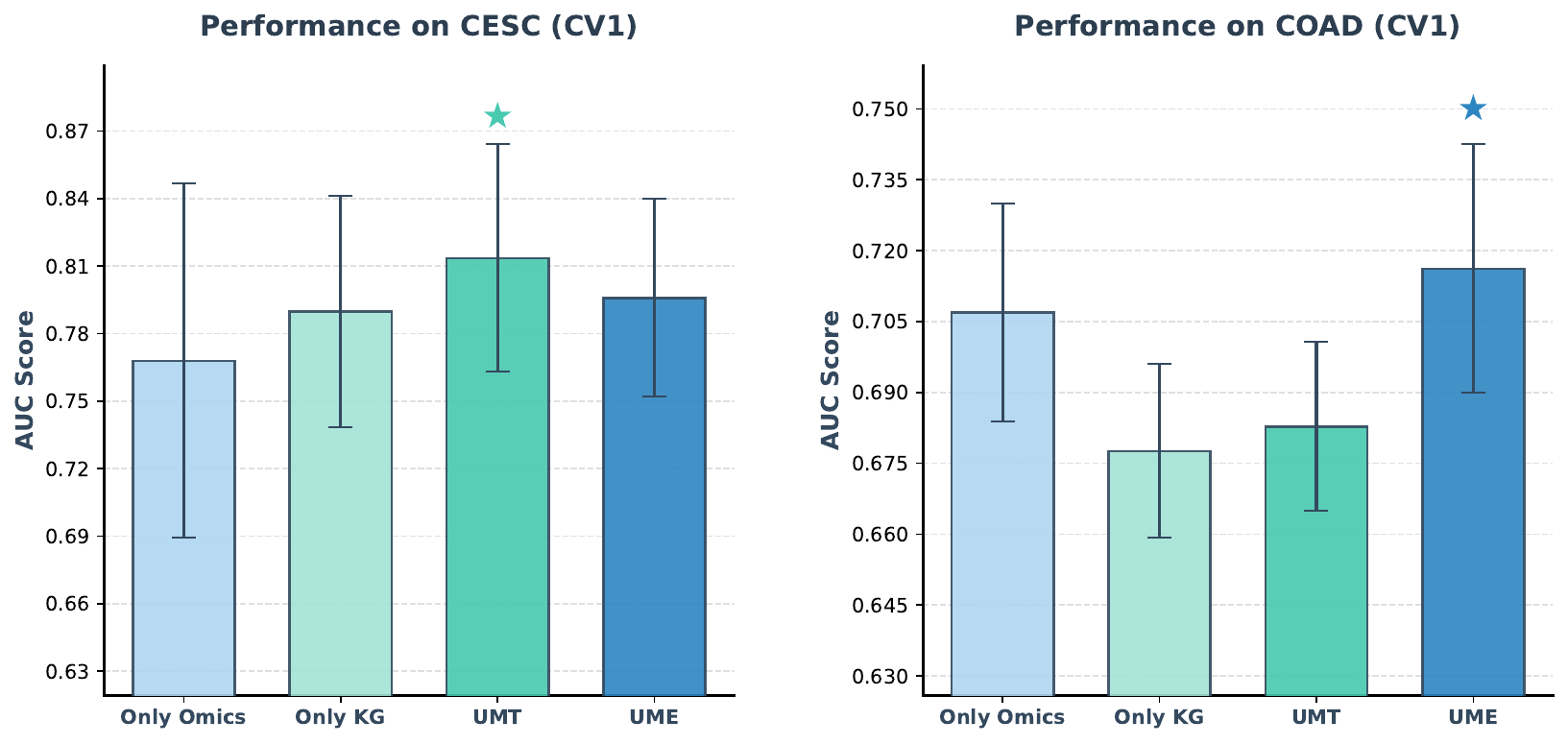}     \caption{
\textbf{Performance comparison of SynLeaF baseline variants
on two cancer datasets.}
This figure shows the AUC scores for the unimodal baseline variants (Only Omics, Only KG) 
and the baseline variants respectively employing two multimodal fusion strategies (UMT, UME)
on the CESC and COAD cancer datasets under the CV1 split.
The height of the bars corresponds to the mean AUC obtained via 5-fold cross-validation, 
while the standard deviations are denoted by the error bars.
The star indicates the optimal fusion strategy,  
which is finally adopted in the SynLeaF adaptive selection mechanism
on that dataset.
    }
    \label{fig:ablation_bar} \end{figure}

In the CESC (CV1) dataset, unimodal Omics and KG yielded AUCs of 0.7679 and 0.7899, respectively. 
A simple late fusion (UME) achieved only 0.7959, struggling to capture deep interactions. 
However, consistent with the first scenario in \citet{Du2023}, 
SynLeaF’s UMT strategy achieved a significant advantage (AUC=0.8136). 
By utilizing feature-level distillation, UMT effectively forces the network to learn from both modalities and retain key cross-modal interactions.
Conversely, COAD (CV1) presents a ``strong omics (0.7069) and weak graph (0.6776)'' scenario. 
Here, forcing feature alignment via UMT introduced noise, dropping the AUC to 0.6828 (lower than Only Omics). 
As proposed by \citet{Du2023} for cases with insignificant paired interactions, 
the UME strategy selected by SynLeaF performed best (AUC=0.7162). 
UME straightforwardly aggregates unimodal results, 
effectively avoiding modality laziness or negative transfer caused by forced cross-modal interactions.

\subsubsection{Parameter Sensitivity Analysis Confirms the Complementary Stability of UMT and UME Strategies}

\begin{figure*}[!htbp]
    \centering
    \includegraphics[width=1\textwidth]{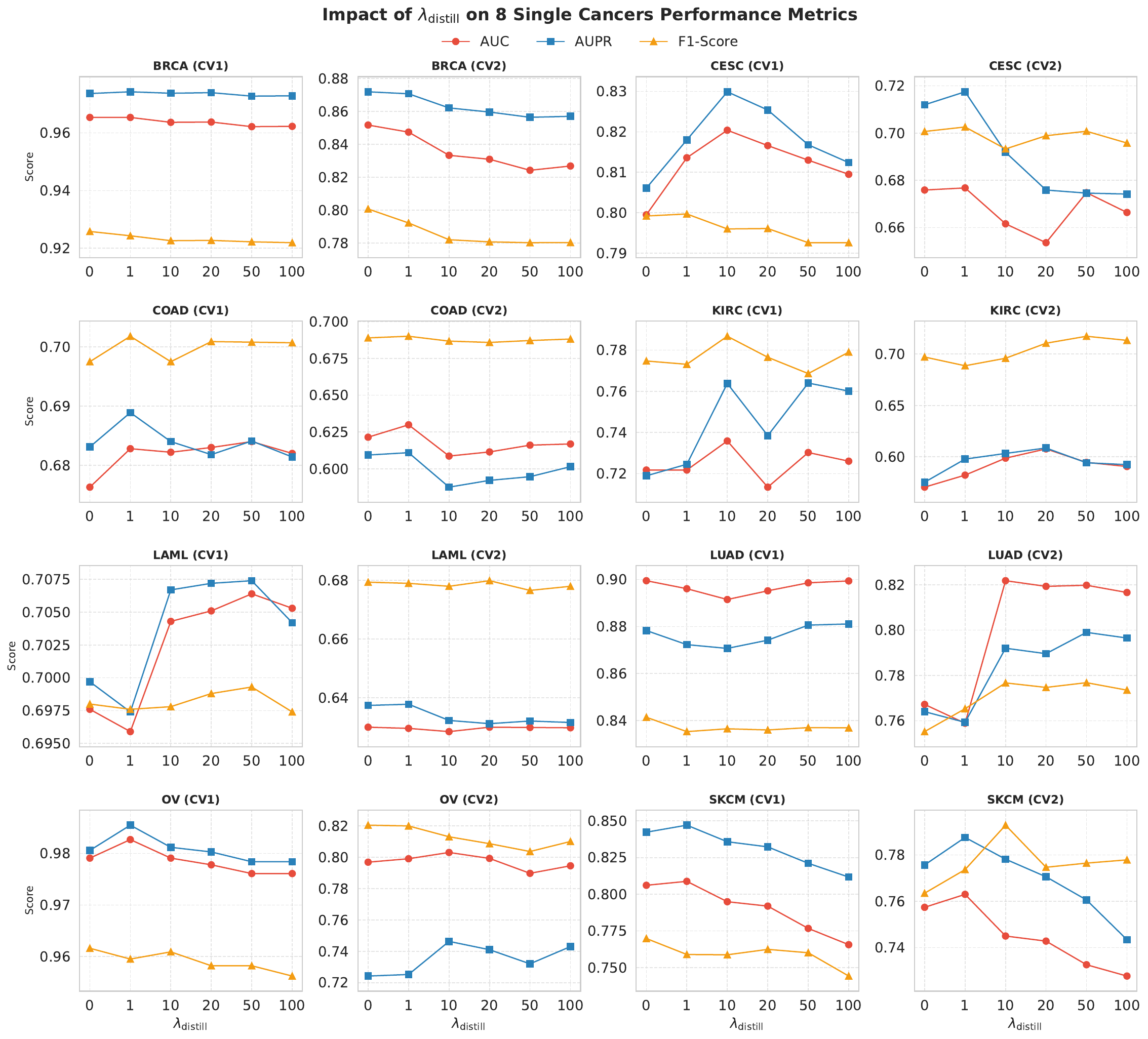} 
    \includegraphics[width=1\textwidth]{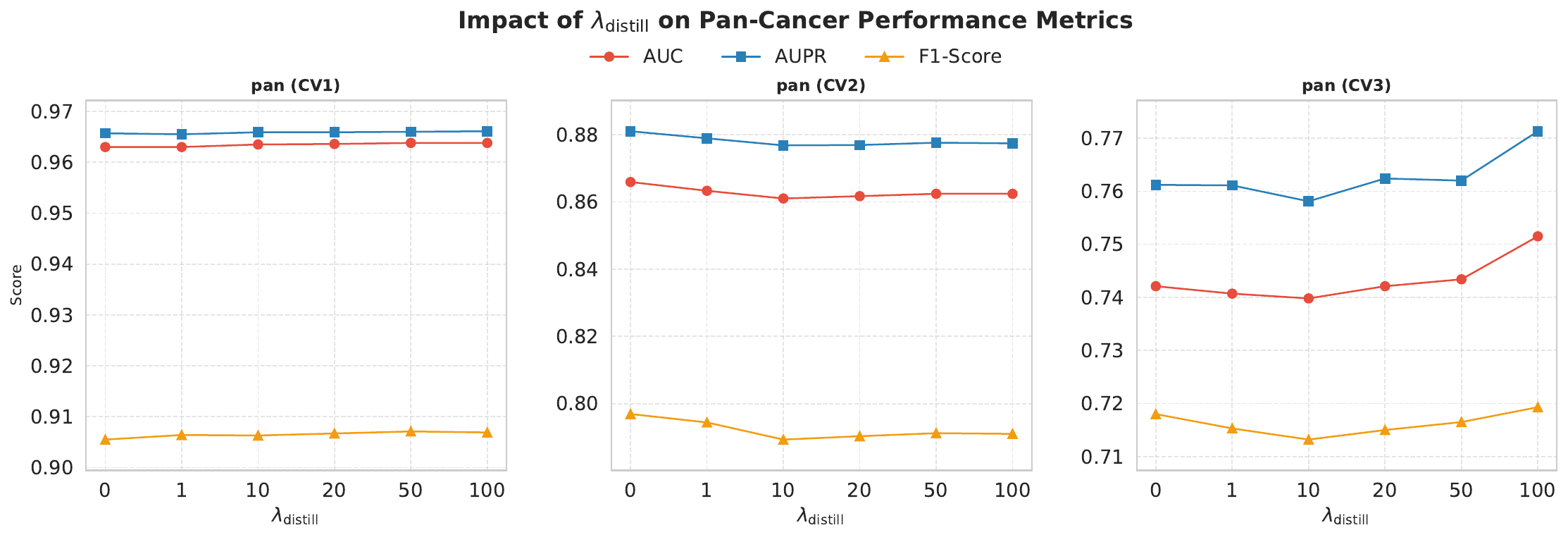} 
    \caption{
\textbf{Sensitivity analysis of the $\lambda_{\text{distill}}$ parameter in the UMT module.}
This figure shows how the performance (AUC, AUPR, F1-Score) of the UMT fusion strategy
changes with the knowledge distillation weight $\lambda_{\text{distill}}$ on all single-cancer and pan-cancer datasets.
A value of $\lambda_{\text{distill}}=0$ corresponds to Na\"ive multimodal training without distillation regularization.
    }
    \label{fig:lambda_sensitivity} 
\end{figure*}

To evaluate the contribution of the feature-level distillation module 
and determine the optimal hyperparameter configuration,
a sensitivity analysis was performed regarding the distillation weight $\lambda_{\text{distill}}$ in equation~\eqref{eqn:loss-umt}
for the UMT module, within the range of $[0, 1, 10, 20, 50, 100]$ (as shown in Figure \ref{fig:lambda_sensitivity}).
Here, $\lambda_{\text{distill}}=0$ is equivalent to Na\"ive early fusion without distillation regularization.

The experimental results show that for most datasets, 
introducing moderate distillation regularization ($\lambda_{\text{distill}}=1$) 
significantly outperforms the no-distillation baseline ($\lambda_{\text{distill}}=0$).
For example, in CESC (CV1), the AUC increased from 0.7995 at $\lambda_{\text{distill}}=0$ to 0.8136 at $\lambda_{\text{distill}}=1$.
This indicates that lightweight feature alignment can effectively mitigate modality laziness 
while avoiding the risk of excessive regularization.
Based on the majority principle and considerations for model generality, 
we uniformly set $\lambda_{\text{distill}}$ to 1 in the final SynLeaF model.

However, this fixed-parameter strategy inevitably faces challenges in some highly heterogeneous datasets.
We observed that LAML (CV1) and LUAD (CV2) showed a special decrease-then-increase trend.
A weak distillation ($\lambda_{\text{distill}}=1$) actually interfered with the model's feature learning, 
leading to performance lower than the baseline with $\lambda_{\text{distill}}=0$.
For example, in LUAD (CV2), the AUC dropped from 0.7672 to 0.7589 when $\lambda_{\text{distill}}=1$.
Although the data trend shows that the model could overcome this obstacle 
and achieve better performance if the distillation weight were further increased (for example, $\lambda_{\text{distill}} \ge 10$),
under the unified setting of $\lambda_{\text{distill}}=1$, the UMT module did indeed reach a performance low.
But during the validation phase, SynLeaF's adaptive strategy successfully 
identified the performance decline of UMT and selected UME as the final inference model.
This allowed the final model to still maintain a competitive performance (the final AUC for LUAD CV2 was 0.8161).
This result precisely validates the necessity of the adaptive selection mechanism 
and the high fault-tolerance of SynLeaF's dual-stage architecture.
It allows the model to use a single set of general hyperparameters for most scenarios, 
while relying on the adaptive switching mechanism to provide a robust fallback 
for the few scenarios that are sensitive to parameters.

\subsection{Gradient Dynamics Analysis Reveals the Mechanism for Mitigating Modality Laziness}

\begin{figure}[!htbp]
    \centering
    \includegraphics[width=1\textwidth]{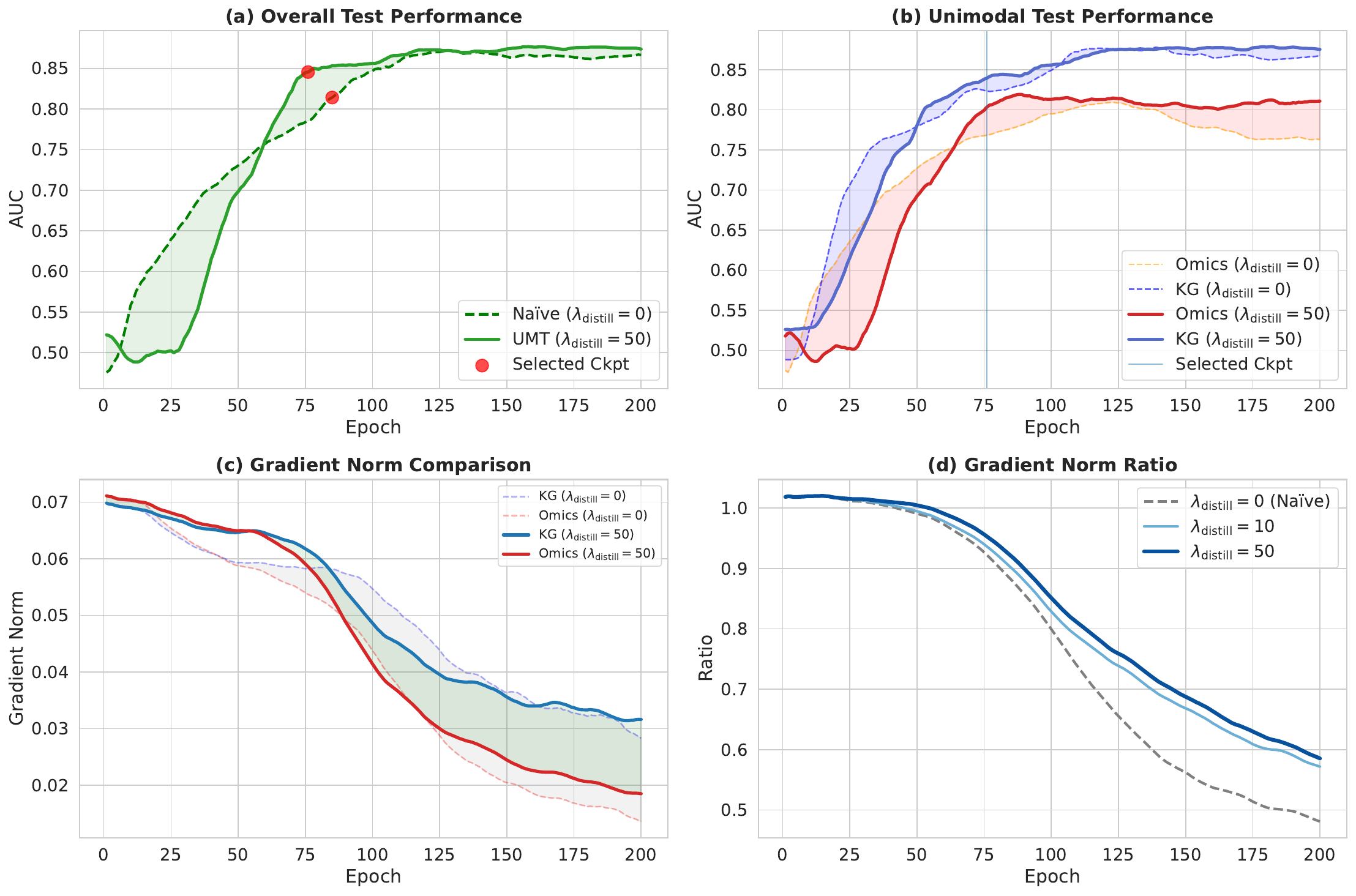}
\caption{
\textbf{Gradient dynamics analysis on the CESC (CV1, Fold2) dataset.}
\textbf{(a)} Overall test performance comparison: Test AUC curves for UMT (solid line) and the Na\"ive no-distillation baseline (dashed line). 
The red dots mark the checkpoints selected based on the validation set.
\textbf{(b)} improvements in unimodal scenarios: change in test AUC for the Omics and KG modalities under the UMT and Na\"ive no-distillation strategies (denoted as Omics/KG ($\lambda_{\text{distill}}=50$) and Omics/KG ($\lambda_{\text{distill}}=0$), respectively). 
The vertical line marks the UMT checkpoint position.
\textbf{(c)} Gradient norm comparison: change in the gradient norms of the two modality encoders
over training epochs under different distillation weights $\lambda_{\text{distill}}$.
\textbf{(d)} Gradient norm ratio: trend of the gradient balance between the two modalities. Here, the gradient norm ratio is defined as the $L_2$ norm of the gradients with respect to the Omics encoder's parameters divided by that of the KG encoder's parameters.
}
    \label{fig:gradient_analysis}
\end{figure}

To better understand how the UMT strategy effectively mitigates the modality laziness problem in multimodal learning,
The CESC (CV1, Fold2) dataset, which shows a typical UMT advantage, 
was employed as a representative case to study the gradient dynamics across the training phase (as shown in Figure \ref{fig:gradient_analysis}).

\subsubsection{The UMT Strategy Significantly Improves Both Overall and Unimodal Performance}

Figure \ref{fig:gradient_analysis}(a) shows the comparison of overall test performance.
In the early stage of training (about the first 50 epochs), the Na\"ive no-distillation baseline converges faster because it lacks additional regularization.
However, the cost of this rapid convergence is overfitting. 
After reaching its peak, the performance fluctuations followed by a decline are observed in the baseline model.
In contrast, although the UMT strategy rises more slowly at the beginning,
it shows a continuous and stable learning ability, and continues to climb after surpassing the baseline around epoch 60,
eventually stabilizing at a high level above 0.86.
The light green shaded area visually illustrates the substantial performance advantage achieved by the UMT strategy during the advanced training epochs.

Figure \ref{fig:gradient_analysis}(b) further reveals that
this overall performance improvement comes from improvements at the unimodal level.
We observed two key phenomena.
First, for the Omics modality, the Na\"ive no-distillation baseline ($\lambda_{\text{distill}}=0$) experienced a significant performance collapse in the later stages of training.
The AUC dropped sharply from its peak to about 0.75, which is a typical phenomenon of overfitting.
In contrast, the UMT strategy ($\lambda_{\text{distill}}=50$) not only increased steadily 
but also successfully avoided performance degradation in the later stages, always staying above 0.80.
Second, for the KG modality, the UMT strategy also maintained a small but consistent advantage compared to the baseline,
indicating that the knowledge distillation mechanism had a positive regularization effect on both modalities.

Notably, the checkpoints denoted by red dots within Figure \ref{fig:gradient_analysis}(a) 
are not the global optimal points of their respective curves on the test set,
and there is still room for improvement after these points.
This observation once again confirms the previous discussion about KIRC, 
showing that the distribution shift between the validation and test sets is a common challenge in biomedical data analysis \cite{subbaswamy2020from}.
Selecting models based on validation set metrics may fail to capture the true optimal state on the test set.

\subsubsection{Gradient Norm Analysis: Knowledge Distillation Enhances the Learning Signal of the Weaker Modality}

To explain the above phenomena from an optimization dynamics perspective,
we analyzed the changes in the gradient norms of the two modal encoders
during the training process (Figure \ref{fig:gradient_analysis}(c)),
following the analytical protocols established in recent multimodal learning studies \cite{peng2022balanced}.

In the Na\"ive no-distillation baseline ($\lambda_{\text{distill}}=0$), we observed a significant asymmetry in gradient decay.
In the early stage of training, the gradient norms of both modalities were 
at a relatively high level of about 0.07.
However, as the training progressed, the gradient norm of the Omics modality dropped sharply, 
eventually falling below about 0.02.
On the other hand, the decrease for the KG modality was relatively gentle, 
finally staying at a level of about 0.03.
This difference means that in the later stages of training, 
the Omics encoder almost stopped learning effectively.
The model mainly relied on the KG modality for prediction, which is direct evidence of modality laziness.

After introducing the knowledge distillation regularization ($\lambda_{\text{distill}}=50$), the situation changed significantly.
The light green shaded area clearly shows how the UMT strategy successfully increased the gradient norm of the Omics modality.
Although the gradient for Omics was still decreasing, the rate of decrease slowed down significantly.
This means that the distillation loss provided an additional supervision signal for the Omics encoder,
forcing it to continue learning to align with the feature representations derived from the pre-trained teacher model and thus avoiding early gradient vanishing.
Notably, under both strategies, the gradients for the KG modality in the later training stages almost overlapped.
This indicates that the main role of UMT is to selectively enhance the weaker modality, 
rather than interfering with the normal learning of the stronger modality.

Figure \ref{fig:gradient_analysis}(d) more directly quantifies the degree of learning balance 
between the two modalities through the gradient ratio (Omics/KG).
A ratio closer to 1.0 indicates that the learning dynamics of the two modalities are more balanced, 
while a decreasing ratio means that the relative contribution of Omics is weakening.
In the Na\"ive no-distillation baseline ($\lambda_{\text{distill}}=0$), the gradient ratio starts from a perfectly balanced 1.0 
and then rapidly decreases, dropping to 0.50 in the later training stages.
This means that by the end of training, 
the gradient contribution of the Omics modality was only half that of the KG modality, 
causing a serious learning imbalance.
As the distillation weight $\lambda_{\text{distill}}$ increases, this imbalance is gradually mitigated.
A value of $\lambda_{\text{distill}}=10$ keeps the ratio at a higher level, and the effect of $\lambda_{\text{distill}}=50$ is the most significant.
Even at the end of training, the ratio remains at about 0.58, 
showing a clear improvement in balance compared to the baseline's value of below 0.50.

The gradient dynamics analysis above provides direct mechanical evidence for the effectiveness of the UMT strategy.
Through the knowledge distillation framework, 
the pre-trained Uni-modal Teacher model provides continuous feature-level supervision for the student encoders in the joint training.
This effectively enhances the gradient signal of the weaker modality and reduces the learning gap with the stronger modality.
We also honestly point out that the UMT strategy did not completely eliminate the gradient gap between modalities. Even with $\lambda_{\text{distill}}=50$,
the final gradient ratio was still 0.58, which is still far from the ideal 1.0.
This suggests that modality laziness is a deep-rooted optimization problem that is difficult to solve completely with a single technique.
However, the experimental results show that 
even mitigating rather than completely eliminating the problem 
is enough to bring substantial improvements in predictive performance,
which validates the rationale of our method's design.

\subsection{SynLeaF Web Server and Case Study Applications}

To facilitate the exploration of synthetic lethality interactions, 
we developed a web-based query platform\footnote{
    The web server is accessible at: \url{https://synleaf.bioinformatics-lilab.cn}
}. 
The backend uses an ensemble of five CV2 models per cancer type 
and averages their outputs to improve robustness.
The interface visualizes predictions across eight cancers and the pan-cancer context,
alongside ground-truth labels from existing datasets for direct comparison.

\begin{table}[!htbp]
\caption{Prediction scores retrieved from the SynLeaF web server for two case studies.}
\label{tab:case_studies}
\tiny
\tabcolsep=0pt
\begin{tabular*}{\textwidth}{@{\extracolsep{\fill}}lccccccccc@{\extracolsep{\fill}}}
\toprule
\textbf{Gene Pairs} & \textbf{BRCA} & \textbf{CESC} & \textbf{COAD} & \textbf{KIRC} & \textbf{LAML} & \textbf{LUAD} & \textbf{OV} & \textbf{SKCM} & \textbf{pan} \\
\midrule
RAD51--BRCA1 & \textbf{0.9928} & 0.3817 & 0.6562 & 0.6038 & 0.5830 & 0.4169 & 0.3849 & 0.5850 & 0.9748 \\
RAD51--BRCA2 & \textbf{0.9213} & 0.4462 & 0.4572 & 0.6160 & 0.5406 & 0.5259 & 0.3624 & 0.6082 & 0.9596 \\
\midrule
ADAR-BRCA1 & \textbf{0.9307} & 0.4876 & 0.5443 & 0.3883 & 0.5413 & 0.4984 & 0.2611 & 0.5221 & 0.8667 \\
ADAR-BRCA2 & \textbf{0.7750} & 0.4697 & 0.3759 & 0.3465 & 0.5174 & 0.4815 & 0.1832 & 0.5595 & 0.7112 \\
\bottomrule
\end{tabular*}
\end{table}

\citet{Milordini2025} demonstrated the synthetic lethality of targeting the RAD51--BRCA2 interaction, 
particularly in pancreatic cancer.
Although these pairs are labeled as positive in our pan-cancer dataset, SynLeaF provides detailed context-specific insights.
As shown in Table \ref{tab:case_studies}, the prediction scores in the BRCA column are exceptionally high,
significantly surpassing those in other cancer types.
This strong signal suggests that breast cancer may share the same RAD51-mediated synthetic lethality mechanism 
observed in pancreatic cancer.
\citet{Chabanon2025} identified ADAR1 as a key synthetic lethal target in BRCA-mutant cancers.
Notably, despite the absence of ADAR-BRCA* pairs from our dataset, SynLeaF successfully predicted this latent relationship, crucially exhibiting high efficacy in the relevant cancer context:
the prediction score in the BRCA column reached 0.9307 for ADAR-BRCA1,
which is remarkably higher than in other unrelated cancer types.
This result demonstrates SynLeaF's capability to generalize and discover novel, 
clinically relevant synthetic lethality pairs in specific cancer types.

\section{Discussion}\label{sec:discussion}

Synthetic lethality is a key mechanism in precision oncology,
yet identifying robust SL pairs remains challenging \cite{ONeil2017}.
Our results show that SynLeaF improves prediction performance 
across pan- and single-cancer settings and adapts to modality dependency differences.
A key finding is that balancing modality interaction 
and avoiding modality laziness are crucial for multimodal SL prediction \cite{Du2023}.
SynLeaF combines unimodal pre-training with two complementary fusion strategies: 
UMT for deep interaction and UME for conservative ensembling, 
providing robustness across diverse cancer contexts and alleviating negative transfer \cite{wang2019characterizing}.

Although SynLeaF has achieved a breakthrough in prediction accuracy, 
the current work
still faces limitations regarding the practical requirements for clinical translation.
First, our model lacks interpretability regarding multimodal interactions. 
Future work will be devoted to clarifying how continuous omics features enhance or suppress KG topology 
to provide a complete biological evidence chain.
Second, the 1:1 class balancing strategy deviates from the highly imbalanced biological reality. 
Furthermore, distribution shifts between validation and test sets 
in small-sample cancers hinder optimal model selection.
Future research could reframe SL prediction as a recommendation task to handle long-tail distributions
and develop more robust evaluation strategies.
Finally, while Cross-VAE handles missing intra-omics data, SynLeaF still requires both omics and KG modalities. 
In clinical practice, a patient may completely lack sequencing data, 
or some new genes may not be included in existing knowledge graphs.
Future studies could explore cross-modal completion techniques 
based on generative modeling to enable flexible prediction 
when only a single fundamental modality is available.

\section{Conclusion}\label{sec:conclusion}

This study introduces SynLeaF, a dual-stage multimodal fusion framework for synthetic lethality prediction across pan- and single-cancer contexts.
Through VAE-enhanced omics encoding and a knowledge distillation strategy, 
SynLeaF effectively overcomes the challenges of modality laziness and heterogeneity in multi-source data fusion.
Extensive experiments show that the proposed framework exhibits state-of-the-art capability within prediction scenarios targeting for both known and new genes.
SynLeaF not only provides a powerful computational tool for discovering synthetic lethality targets,
but its approach to handling dynamic multimodal dependencies also 
offers a new perspective for general link prediction tasks in the biomedical field.

\section{Methods}\label{sec11}

\subsection{Data Acquisition and Preprocessing}

We constructed a strictly filtered and multi-source integrated pan-cancer and cancer-specific dataset.
The data covers SL gene pairs, multi-omics data, biomedical knowledge graphs, and protein sequence information.
We integrated data from authoritative databases 
such as SynLethDB 2.0 \cite{Wang2022}, ELISL \cite{Tepeli2023}, TCGA \cite{Cerami2012}, and UniProt \cite{citeniProt},
and we designed a standardized preprocessing workflow to eliminate noise and heterogeneity.

\subsubsection{Data Sources}

\textbf{Synthetic Lethality Data.} 
The synthetic lethality label data comes from two main sources.
For the pan-cancer prediction task, we use the SynLethDB 2.0 database,
which provides general synthetic lethality gene pairs across cancer types.
For the cancer-specific prediction task, we adopted the dataset organized by the ELISL study.
ELISL integrates high-confidence SL gene pairs 
from previous studies such as DiscoverSL \cite{Das2019}, ISLE \cite{lee2018harnessing}, and EXP2SL \cite{Wan2020}.
It covers eight specific cancer types,
and the number of synthetic lethality pairs before preprocessing is shown in Supplementary Table S3.

\textbf{Multi-Omics Data.} 
The omics data comes from the TCGA database curated by the cBioPortal platform \cite{cerami2012cbio}.
We collected four key omics features for the above eight cancer types: 
Copy Number Variation, which represents the relative linear copy number variation values of genes; 
Gene Expression, which uses mRNA z-score data processed by RNA-Seq V2 RSEM normalization; 
DNA Methylation, which reflects the epigenetic regulation status of genes (HM27 or HM450); 
and Mutation data, which records the mutation status of each gene in every sample, 
where we only consider the mutation counts of genes in each sample.
For the pan-cancer dataset, due to data limitations, 
the dataset is restricted to three data modalities consisting of gene expression, mutation, and copy number variation.

\textbf{Knowledge Graph.} 
We adopted the biomedical knowledge graph (SLKG 2.0) provided by SynLethDB 2.0.
This knowledge graph contains 37,341 entities of 11 types (including genes, Gene Ontology, pathways, drugs, and diseases) 
and 1,405,652 relationships of 27 types, forming a rich biomedical knowledge network.

\textbf{Protein Sequences.} 
To support comparative experiments, we downloaded reviewed human protein sequence data from the UniProt database (as of October 26, 2024), 
which contains a total of 20,428 sequences where each gene corresponds to a representative protein sequence consisting of 21 types of amino acids.

\subsubsection{Data Preprocessing}

We executed rigorous protocols for gene filtering and data cleaning to guarantee the 
uniformity and superior quality of the multimodal data.

\textbf{Gene Filtering and Alignment.} 
We performed multi-condition intersection filtering on the genes.
Genes included in the study must meet the following conditions simultaneously: 
they must have corresponding reviewed protein sequences in UniProt, 
be recorded in at least one type of omics data, 
and have at least one accessible neighbor node in the knowledge graph.
Through this strict filtering mechanism, we filtered out genes with incomplete information 
and ensured that every gene used for model training has a complete multimodal feature representation.
The number of synthetic lethality pairs after filtering is shown in Table \ref{tab:sl_count_after}.

\begin{table}[!htbp]
    \centering
        \caption{Statistics of synthetic lethality pair counts after data processing}
    \label{tab:sl_count_after}     \begin{tabular}{lccc}
        \toprule
Cancer Type & \# Total Genes & \# Pos. & \# Neg. \\
        \midrule
BRCA	& 17965 &	1349	& 990 \\
CESC	& 17977 &	144	    & 4738 \\
COAD	& 17961 &	1560	& 70982 \\
KIRC	& 17963 &	60	    & 2514 \\
LAML	& 17944 &	1147	& 18912 \\
LUAD	& 17954 &	582	    & 5460 \\
OV	    & 17958 &	253	    & 556 \\
SKCM	& 17969 &	101	    & 16157 \\
pan 	& 17690 &	33746 	& 3509 \\
        \bottomrule
    \end{tabular}
\footnotetext{\# Pos. denotes the count of positive samples; \# Neg. denotes the count of negative samples.}
\end{table}

\textbf{SL Data Construction and Balancing.} 
To address the noise and imbalance issues in the original SL data, we performed the following processing.
First, in the pan-cancer dataset, we treated pairs explicitly labeled as Non-SL and Synthetic Rescue as negative samples.
Second, for conflicting gene pairs that appeared in both the positive and negative sample sets, 
we adopted a conservative strategy and uniformly classified them as negative samples.
Additionally, we corrected errors in the ELISL data 
where some self-loops were incorrectly identified as positive samples.
Consequently, we reassigned them to the negative category.
Considering the inherent sparsity of genuine synthetic lethality associations,
the quantity of negative samples within the dataset typically vastly surpasses the count of positive ones.
To avoid the model biasing towards the majority class, 
we adopted a 1:1 positive-negative sample balance strategy.
When there were too many negative samples, we used random undersampling; 
when known negative samples were insufficient in specific cancer types, 
we randomly generated unlabeled gene pairs from the filtered gene pool as supplementary negative samples, 
and we ensured that these generated pairs did not overlap with known positive samples.

\begin{figure*}[!htbp]
    \centering
    \includegraphics[width=1\textwidth]{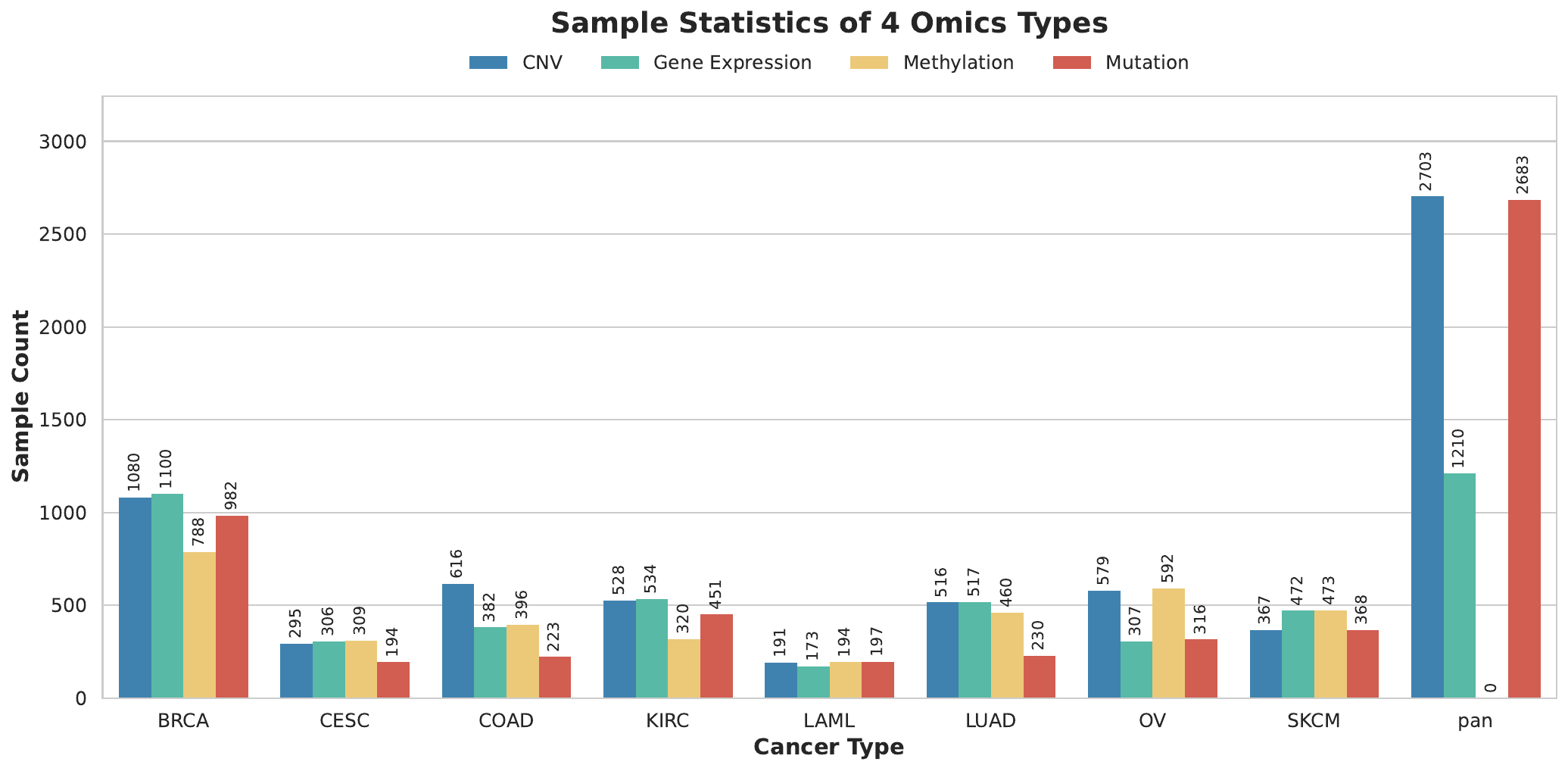}     \caption{
\textbf{Statistics of sample counts for original omics data across cancer types.}
This figure displays the number of raw available samples 
before preprocessing for four omics data types 
(Copy Number Variation (CNV), Gene Expression, DNA Methylation, and Mutation) 
across the eight specific cancer types and the pan-cancer dataset used in this study.
Noting that the pan-cancer (pan) dataset does not include DNA methylation data due to data source limitations.
    }
    \label{fig:omics_sample_stat} \end{figure*}

\textbf{Omics Data Normalization.} 
The sample counts for different features across various cancer types in the original omics data are shown in Figure \ref{fig:omics_sample_stat}.
We performed sample-level alignment on the omics data for each cancer 
to ensure that the four omics features for each case sample are one-to-one matched.

To address the sparsity and distribution characteristics of the data, the processing workflow is as follows.
Due to the sparsity of omics data, we filled all missing values and non-numeric values with 0 
to represent no abnormality or a default state;
for gene expression data, we truncated values smaller than -10 to -10 
because we observed that extremely small values could cause numerical instability;
regarding mutation profiles, we calculated the mutation frequency for every individual gene 
and applied a logarithmic transformation to map it to the interval $[0, 1]$ using the following formula:

\begin{equation}
x' = \begin{cases}
\frac{\ln(1+x)}{\ln(1+M)}, & \text{if } M > 0 \\
0, & \text{if } M = 0
\end{cases}
\end{equation}

where $x$ is the original mutation count 
and $M$ is the maximum mutation count in that cancer type.
Furthermore, given that synthetic lethality effects often involve interactions 
between normal genes and abnormal genes \cite{kaelin2005concept},
we did not exclude genes with low variation 
based on the case sample distribution 
but instead retained all genes that met the omics alignment requirements.
We retained only unique records for duplicate gene entries appearing in single-omics data.

\textbf{Knowledge Graph Refinement.} 
To construct cancer-specific knowledge graphs, 
we performed subgraph extraction on the original graph based on cancer types.
For each specific cancer, we only retained its corresponding ``Disease'' node and the edges directly connected to it, 
removing other irrelevant cancer nodes to reduce noise interference.
Additionally, for the unidirectional relationships in the graph, 
we generated corresponding reversed edges to transform the directed graph into a heterogeneous graph 
containing bidirectional information.

\subsubsection{Dataset Splitting Strategies}

In order to thoroughly assess the robustness and generalization capacity of our framework, 
we referred to the work of PiLSL \cite{Liu2022} and adopted three cross-validation (CV) splitting strategies.
The partitioning of all datasets adhered to a fixed proportion of 7:1:2 for the training, validation, and testing subsets, respectively.

\begin{itemize}
    \item \textbf{CV1 (Random Split)}: 
        All samples (gene pairs) are randomly shuffled and split. 
                Under this setting, genes present in the testing phase might also exist within the training set.
    This primarily evaluates the inductive potential of the model concerning novel combinations of known genes.
    
    \item \textbf{CV2 (Semi-New Gene Split)}: 
                    The gene set is partitioned to guarantee that for every gene pair situated in the validation or testing subsets, 
    precisely a single gene is found in the training data,   
    while the other gene is completely new to the training set.
        This simulates the real-world scenario of finding known targets for new genes, 
    and it evaluates the model's generalization ability for semi-new gene pairs.

    \item \textbf{CV3 (New Gene Split)}: 
        This is the strictest splitting method, 
        ensuring a complete absence of test-set genes within the training set. 
        This tests the model's inferential capability in a completely unexplored gene space (zero-shot setting).
\end{itemize}

It is worth noting that due to the small sample size of some cancer-specific datasets, 
performing CV3 splitting may result in extremely few samples in the test set, rendering it statistically insignificant.
Therefore, in subsequent experiments, we only report the results of CV1 and CV2 for single-cancer prediction tasks, 
while we use all three splitting strategies for comprehensive evaluation in the pan-cancer prediction task.

\subsection{The SynLeaF Framework}

This paper introduces the SynLeaF architecture, a deep learning system grounded in dual-stage multimodal fusion strategies.
This framework aims to accurately predict SL interactions 
by integrating omics features from genomics and structured knowledge from biomedical KGs.
The architecture of the model is illustrated in Figure \ref{fig:model_architecture}.

\subsubsection{Problem Definition}

We define the synthetic lethality prediction task as a binary classification problem.
Given a specific instance of a gene pair $(g_i, g_j)$, the objective is to predict
whether an SL relationship exists between them, 
which corresponds to outputting a label $y \in \{0, 1\}$.
Our method follows a unified Siamese-like network architecture \cite{chicco2021siamese} for representation learning of gene pairs, 
regardless of whether it is in a single-modal or multimodal setting.
Specifically, for each gene in the pair, 
we first utilize its multimodal features to learn its embedding representation 
through a corresponding modality-specific encoder $\varphi_m$ (where $m \in \{o, k\}$ represents the omics or knowledge graph modality):

\begin{equation}
\mathbf{h}_i^m = \varphi_m(\mathbf{f}_i^m), \quad \mathbf{h}_j^m = \varphi_m(\mathbf{f}_j^m)
\end{equation}
where $\mathbf{f}_i^m$ denotes the feature input corresponding to the $m$-th modality for the gene $g_i$.

The obtained embedding vectors of the two genes are subsequently used as input features 
to output the final prediction logits through a shared classifier $\mathcal{F}_m$:
\begin{equation}
\hat{y}_{i,j}^m = \mathcal{F}_m([\mathbf{h}_i^m; \mathbf{h}_j^m])
\end{equation}
where $[\cdot;\cdot]$ denotes the concatenation operation.

We aim to optimize the model by minimizing the Binary Cross-Entropy (BCE) loss, which is formulated as:

\begin{equation}
\mathcal{L}_{\text{BCE}} = - \frac{1}{|\mathcal{D}|} \sum_{(g_i, g_j) \in \mathcal{D}} \left( y_{i,j} \log(\varsigma(\hat{y}_{i,j}^m)) + (1 - y_{i,j}) \log(1 - \varsigma(\hat{y}_{i,j}^m)) \right)
\label{eq:bce}
\end{equation}
where $\mathcal{D}$ represents the set of all gene pairs within the training set, 
$y_{i,j}$ denotes the ground truth annotation,
and $\varsigma(\cdot)$ serves as the sigmoid activation function for transforming logits into probability scores.

We denote the predicted probability after sigmoid transformation as
$p_{i,j}^m = \varsigma(\hat{y}_{i,j}^m)$, 
which represents the probability that the gene pair $(g_i, g_j)$ is predicted to have a synthetic lethality relationship under modality $m$.

\subsubsection{Omics Encoder with VAE Early Fusion}

The encoding of the omics modality adopts an early fusion module (OmicsEncoder, $\varphi_{o}$).
We constructed an $N \times N$ VAE encoder matrix 
(where $N=4$ denotes the total count of omics categories)
and leveraged the Product of Experts (PoE) \cite{Kutuzova2021} mechanism 
to execute early fusion on the omics features.
Each VAE encoder is implemented by a Multi-Layer Perceptron (MLP), 
which maps input features to the mean $\boldsymbol{\mu}$ and log-variance $\log\boldsymbol{\sigma}^2$ in the latent space.

Let the $N$ types of omics features for gene $g_i$ be denoted as $\mathbf{F}_i = \{\mathbf{f}_i^{(1)}, \dots, \mathbf{f}_i^{(N)}\}$.
As shown in the Omics Encoder module of Figure \ref{fig:model_architecture},
the encoder matrix consists of self-encoders (Self-VAE) on the diagonal and cross-encoders (Cross-VAE) off the diagonal.
For the $k$-th omics data $\mathbf{f}_i^{(k)}$ of gene $i$, 
the self-encoder $\text{VAE}_{k,k}$ maps it to the parameters of the latent distribution, $\boldsymbol{\mu}_{i}^{(k, \text{self})}$ and $\log(\boldsymbol{\sigma}_{i}^{(k, \text{self})})^2$.
For the $j$-th ($j \neq k$) omics data $\mathbf{f}_i^{(j)}$ of gene $i$, 
the cross-encoder $\text{VAE}_{k,j}$ attempts to infer the latent distribution of the $k$-th omics type, 
with parameters denoted as $\boldsymbol{\mu}_{i}^{(k, \text{cross}, j)}$ and $\log(\boldsymbol{\sigma}_{i}^{(k, \text{cross}, j)})^2$.

To aggregate features from different perspectives 
and address the deficiency of missing modalities for certain genes, 
we utilize the PoE mechanism to calculate a joint posterior distribution for each target omics type $k$ of gene $i$, 
where the parameters $(\boldsymbol{\mu}_{i}^{(k, \text{PoE})}, \boldsymbol{\sigma}_{i}^{(k, \text{PoE})})$ are given by the following formulas:

\begin{equation}
\frac{1}{(\boldsymbol{\sigma}_{i}^{(k, \text{PoE})})^2} = \sum_{j \neq k} \frac{1}{(\boldsymbol{\sigma}_{i}^{(k, \text{cross}, j)})^2}, \quad
\boldsymbol{\mu}_{i}^{(k, \text{PoE})} = (\boldsymbol{\sigma}_{i}^{(k, \text{PoE})})^2 \sum_{j \neq k} \frac{\boldsymbol{\mu}_{i}^{(k, \text{cross}, j)}}{(\boldsymbol{\sigma}_{i}^{(k, \text{cross}, j)})^2}
\end{equation}

Then, we sample from the posterior distributions of the self-encoder path and the PoE path 
using the reparameterization trick to obtain the respective sets of latent variables
$\mathbf{Z}_i^{\text{self}} = \{\mathbf{z}_i^{(k, \text{self})}\}_{k=1}^{N}$ and $\mathbf{Z}_i^{\text{PoE}} = \{\mathbf{z}_i^{(k, \text{PoE})}\}_{k=1}^{N}$:

\begin{equation}
\mathbf{z}_i^{(k, \text{path})} = \boldsymbol{\mu}_i^{(k, \text{path})} + \boldsymbol{\sigma}_i^{(k, \text{path})} \odot \boldsymbol{\epsilon}, \quad \text{path} \in \{\text{self}, \text{PoE}\}
\label{eq:reparam}
\end{equation}
where $\boldsymbol{\epsilon} \sim \mathcal{N}(0, \mathbf{I})$ denotes noise sampled from a standard normal distribution.

Finally, the latent variable sets from the two paths 
are subjected to mean pooling respectively and then concatenated.
They are then projected via a fully connected layer to yield the final omics embedding $\mathbf{h}_i^{o}$:

\begin{equation}
\mathbf{h}_i^{o} = \text{FC}\left(\left[\frac{1}{N}\sum_{k=1}^{N} \mathbf{z}_i^{(k, \text{self})}; \frac{1}{N}\sum_{k=1}^{N} \mathbf{z}_i^{(k, \text{PoE})}\right]\right)
\label{eq:omics_fusion}
\end{equation}

The training process adopts a Variational Information Bottleneck (VIB) loss function \cite{alemi2017deep},
which includes reconstruction error and a KL divergence regularization term:
\begin{equation}
\mathcal{L}_{\text{omics}} = \mathcal{L}_{\text{BCE}} + \lambda_{\text{self}} \sum_{k} D_{KL}(q_{\text{self}}^{(k)} || p) + \lambda_{\text{cross}} \sum_{k} D_{KL}(q_{\text{PoE}}^{(k)} || p)
\end{equation}
where $p$ is the standard normal prior distribution $\mathcal{N}(0, \mathbf{I})$. 
The weights of the two KL divergences, $\lambda_{\text{self}}$ and $\lambda_{\text{cross}}$, are set to 0.1 and 0.5, respectively.

\subsubsection{Knowledge Graph Encoder with RGCN}

For the KG modality, we adopt RGCN \cite{Schlichtkrull2017} as the encoder (KGEncoder, $\varphi_k$)
to capture the topological structures and semantic relationships of genes in the biological network.

Let $\mathcal{G} = (\mathcal{V}, \mathcal{E}, \mathcal{R})$ be a knowledge graph, 
comprising the collections of entities $\mathcal{V}$, edges $\mathcal{E}$, and relations $\mathcal{R}$.
Regarding a specific target gene $g_i$ (corresponding to graph node $i$), 
we first employ an $L$-hop subgraph sampling strategy to extract its local neighborhood subgraph to reduce computational complexity.
The initial node features are obtained through an Embedding Layer.
The RGCN layer updates node representations by aggregating neighbor information under different relationship types $r \in \mathcal{R}$:

\begin{equation}
    \mathbf{h}_i^{(l+1)} = \mathrm{ReLU}\left(\sum_{r \in \mathcal{R}} \sum_{j \in \mathcal{N}_i^r} \frac{1}{c_{i,r}} \mathbf{W}_r^{(l)} \mathbf{h}_j^{(l)} + \mathbf{W}_0^{(l)} \mathbf{h}_i^{(l)}\right)
\end{equation}
where $\mathcal{N}_i^r$ constitutes the neighbor set for node $i$ associated with relationship $r$,
while $c_{i,r}$ serves as a normalization factor equal to $|\mathcal{N}_i^r|$.

After multi-layer RGCN aggregation, we extract the final representation $\mathbf{h}_i^{k}$ of the central gene node.
Optimization of the knowledge graph branch is likewise achieved 
through the minimization of the binary cross-entropy loss, 
denoted as $\mathcal{L}_{\text{kg}} = \mathcal{L}_{\text{BCE}}$.

\subsubsection{Dual-Stage Multimodal Fusion Strategy}
\label{sec:models}

The multimodal fusion adopts a dual-stage training strategy, which includes two complementary schemes named UMT and UME \cite{Du2023}.
In the first stage, we independently train the omics prediction model (Only Omics) and the knowledge graph prediction model (Only KG).
Let the omics teacher encoder obtained from the first stage training be $\varphi_o^T$, and the knowledge graph teacher encoder be $\varphi_k^T$.
This stage ensures that each single-modal encoder can fully explore the feature representations within its modality without interference from other modalities.
In the second stage, we designed two complementary fusion strategies and dynamically identified the optimal scheme according to the results observed in the validation subset.

(1) \textbf{Uni-Modal Teacher (UMT)}: 
This strategy employs a Knowledge Distillation framework to address the issue of modality laziness 
and guarantee that the multimodal architecture comprehensively captures the unimodal feature representations specific to each modality.
We freeze the single-modal encoders pre-trained in the first stage as the ``Teacher'' 
and initialize a new multimodal model as the ``Student'', 
which includes an omics student encoder $\varphi_o^S$ and a knowledge graph student encoder $\varphi_k^S$.
While learning the classification task, the student model is also required to 
simulate the intermediate feature representations of the teacher model 
by minimizing the distillation loss (Mean Squared Error, MSE).
The overall loss function is formulated as follows:

\begin{equation}
\label{eqn:loss-umt}
\mathcal{L}_{\text{UMT}} = \mathcal{L}_{\text{BCE}} + \lambda_{\text{distill}} \sum_{m \in \{o, k\}} \|\mathbf{h}_i^{m, S} - \mathbf{h}_i^{m, T}\|^2 + \mathcal{L}_{\text{KL}}
\end{equation}
Here, $\mathbf{h}_i^{o, T}$ and $\mathbf{h}_i^{k, T}$ are the omics and graph embeddings 
output by the frozen teacher encoders for gene $i$, respectively.
While $\mathbf{h}_i^{o, S}$ and $\mathbf{h}_i^{k, S}$ are the corresponding outputs of the student encoders.

The classifier $\mathcal{F}_{\text{UMT}}$ receives the genomic and knowledge graph features 
$[\mathbf{h}_i^{o, S}; \mathbf{h}_i^{k, S}]$ 
extracted by the student encoders to make predictions.
The hyperparameter $\lambda_{\text{distill}}$ is set to 1, 
and this feature-level distillation forces each branch of the student model to maintain strong feature extraction capabilities,
which effectively alleviates modality laziness.

(2) \textbf{Uni-Modal Ensemble (UME)}: 
As a simpler late fusion baseline, UME directly integrates the prediction results of the two pre-trained models from the first stage.
The final predicted probability $p_{\text{UME}}$ is the average of the output probabilities from the two single-modal models:

\begin{equation}
p_{\text{UME}} = \frac{1}{2} \left(\varsigma(\hat{y}_{o}) + \varsigma(\hat{y}_{k})\right)
\end{equation}

UME requires no additional training and completely avoids gradient interference issues during joint training, 
making it more robust in cases where there are significant differences between modalities.

Considering the differences in the effects of cross-modal interactions across different datasets, 
we adopted a data-driven adaptive selection strategy.
After training is completed, we evaluate the AUC metrics of UMT and UME on the validation set separately 
and select the better-performing method as the final model:

\begin{equation}
p_{i,j} = 
\begin{cases}
p_{i,j}^{\text{UMT}}, & \text{if } \text{AUC}_{\text{UMT}}^{\text{val}} > \text{AUC}_{\text{UME}}^{\text{val}} \\
p_{i,j}^{\text{UME}}, & \text{otherwise}
\end{cases}
\end{equation}

UMT typically performs better when both omics and knowledge graph single-modal features are strong 
and their interaction provides additional information.
Conversely, when one modality is significantly dominant or when cross-modal interaction introduces noise, 
the simple ensemble strategy of UME proves to be more effective.

It is worth noting that compared to traditional complex fusion methods, 
our method is simple to implement and easy to tune.
UMT requires tuning only one additional hyperparameter (the distillation loss weight $\lambda_{\text{distill}}$), 
while UME does not even require extra training.
This simplicity not only improves the practicality of the method but also enhances its portability to new datasets.

\backmatter

\section*{Code availability}

The data and code can be accessed at the following GitHub repository: \url{https://github.com/Jmpax404/SynLeaF}.

\section*{Acknowledgements}
This work was supported by the grants from the National Key R\&D Program of China (2024YFA0919600) and National Natural Science Foundation of China (32470704).

\section*{Author contributions}
J.L. conceived and designed the project and supervised the work. Z.X. developed the methods, performed bioinformatics analysis and drafted the manuscript. S.Z., R.W., and R.H. prepared the data and performed the benchmarks. S.Z., S.C., Y.H., and J.M. contributed to the benchmarks. Y.C., X.W., and Y.W. participated in project design and coordination.

\section*{Competing interests}

The authors declare no competing interests.


\end{document}


\title[Supplementary Information for\\ SynLeaF: A Dual-Stage Multimodal Fusion Framework for Synthetic Lethality Prediction Across Pan- and Single-Cancer Contexts]{Supplementary Information for\\ SynLeaF: A Dual-Stage Multimodal Fusion Framework for Synthetic Lethality Prediction Across Pan- and Single-Cancer Contexts}

\author[1]{\fnm{Zheming} \sur{Xing}}

\author[1]{\fnm{Siyuan} \sur{Zhou}}

\author[1]{\fnm{Ruinan} \sur{Wang}}

\author[1]{\fnm{Rui} \sur{Han}}

\author[1]{\fnm{Shiming} \sur{Zhang}}

\author[1]{\fnm{Shiqu} \sur{Chen}}

\author[1]{\fnm{Yurui} \sur{Huang}}

\author[3]{\fnm{Jiahao} \sur{Ma}}

\author[4]{\fnm{Yifan} \sur{Chen}}

\author[1]{\fnm{Xuan} \sur{Wang}}

\author[2]{\fnm{Yadong} \sur{Wang}}

\author*[1,2]{\fnm{Junyi} \sur{Li}}\email{lijunyi@hit.edu.cn}

\affil*[1]{\orgdiv{School of Computer Science and Technology}, \orgname{Harbin Institute of Technology (Shenzhen)}, \orgaddress{ \city{Shenzhen},  \state{Guang Dong} \postcode{518055}, \country{China}}}

\affil[2]{\orgdiv{Key Laboratory of Biological Bigdata, Ministry of Education}, \orgname{Harbin Institute of Technology}, \orgaddress{ \city{Harbin}, \state{Heilongjiang} \postcode{150001}, \country{China}}}

\affil[3]{\orgdiv{School of Biomedical Sciences}, \orgname{The University of Hong Kong}, \orgaddress{\state{Hong Kong SAR}, \country{China}}}

\affil[4]{\orgdiv{Departments of Mathematics and Computer Science}, \orgname{Hong Kong Baptist University}, \orgaddress{\state{Hong Kong SAR}, \country{China}}}

\maketitle

\section{Effectiveness Validation of the Unimodal Encoder Design}

To better understand the contribution of each component in SynLeaF and its behavior in different biological scenarios,
we conducted extensive unimodal analysis and ablation experiments 
utilizing the pan-cancer dataset alongside all eight cancer-specific datasets under the CV1 split.
In this section, we take specific cancers as representatives for detailed analysis 
to reveal the common patterns and specific phenomena behind the model design.

\subsection{Complementarity and Dominance of Omics Data}

To investigate in-depth the working mechanism of the SynLeaF omics early fusion module,
and to validate the necessity of integrating multi-source omics data, 
we conducted leave-one-out ablation experiments across all datasets under the CV1 split.
Here, we take COAD and SKCM cancers as representatives for detailed analysis.
As presented in Supplementary Table \ref{tab:omics_ablation_transposed}, 
through a comparative analysis of the complete omics model with variants missing a single modality, 
we revealed the very different omics dependency patterns in different cancer types.

\begin{table}[!htbp]
\centering
\caption{Ablation experiment results of the leave-one-out strategy for omics data across pan-cancer and 8 specific cancers under CV1 (Metric: AUC)}
\label{tab:omics_ablation_transposed}
\begin{tabular*}{\textwidth}{@{\extracolsep{\fill}}lccccc@{\extracolsep{\fill}}}
\toprule
Cancer & Full Omics & w/o CNV & w/o Gene Exp. & w/o Methylation & w/o Mutation \\
\midrule
BRCA & \textbf{0.9663} & 0.9651 & 0.9638 & 0.9643 & 0.9649 \\
CESC & 0.7679 & 0.7432 & 0.7357 & \textbf{0.7951} & 0.7784 \\
COAD & \textbf{0.7069} & 0.6949 & 0.6555 & 0.6892 & 0.6950 \\
KIRC & 0.6505 & \textbf{0.6968} & 0.6106 & 0.6591 & 0.6690 \\
LAML & 0.6838 & 0.6860 & 0.6758 & 0.6873 & \textbf{0.6922} \\
LUAD & 0.8794 & \textbf{0.8941} & 0.8810 & 0.8881 & 0.8852 \\
OV & \textbf{0.9819} & 0.9760 & 0.9755 & 0.9778 & 0.9808 \\
SKCM & 0.6390 & \textbf{0.6568} & 0.5907 & 0.6172 & 0.5262 \\
pan & \textbf{0.9631} & 0.9621 & 0.9629 & - & 0.9555 \\
\bottomrule
\end{tabular*}
\end{table}

In the COAD dataset, the complete multimodal version of SynLeaF achieved the best performance (AUC = 0.7069).
Removing any single type of omics data led to a performance decrease, 
with the removal of gene expression data (w/o Gene Exp.) having the most significant impact, causing the AUC to drop to 0.6555.
This indicates that in synthetic lethality prediction for colon cancer, 
transcriptome data provides the most important information, 
but mutation, methylation, and copy number variation data also offer essential complementary perspectives.
The PoE mechanism of SynLeaF successfully integrates these complementary weak signals into a stronger predictive signal.

In the SKCM dataset, we observed an extreme modality dependency.
Removing the mutation data (w/o Mutation) caused the AUC to drop dramatically from 0.6390 to 0.5262.
This sharp drop proves that mutation information plays an absolutely dominant role in skin cutaneous melanoma.
Interestingly, the experimental data show that removing the CNV data (w/o CNV) actually increased the AUC to 0.6568,
which suggests that under the current data processing method, 
the copy number variation data for SKCM might contain noise that interferes with the prediction.
Although the inclusion of CNV introduced some noise, SynLeaF's Full Omics model (0.6390) still maintained a robust performance,
which was far superior to the case where the key modality was lost.
This result demonstrates that the VAE-based fusion module has good noise robustness.
Even if the input contains some redundant or noisy modalities, 
the model can still automatically focus on the high-value modalities, 
thus avoiding the fussy process of manual feature selection.

\subsection{Parameter Stability of Cross-VAE}

To investigate how the VAE module in SynLeaF balances feature reconstruction and latent distribution alignment, 
we performed a sensitivity analysis regarding the weight assigned to the KL divergence.
The total loss function for the VAE comprises the reconstruction error and the KL divergence,
where $\lambda_{\text{self}}$ controls the distribution regularization within a single modality,
and $\lambda_{\text{cross}}$ controls the cross-modal consistency constraint through the PoE mechanism.
We performed a grid search over the ranges 
of $\lambda_{\text{self}} \in \{0.01, 0.1, 1\}$ and $\lambda_{\text{cross}} \in \{0.1, 0.5, 1\}$ across all datasets. 
Taking COAD and OV cancers as representatives for analysis, 
the results illustrated in Supplementary Figure \ref{fig:kl_heatmap} demonstrate the very high parameter robustness of SynLeaF.

\begin{figure}[!htbp]
    \centering
    \includegraphics[width=1\textwidth]{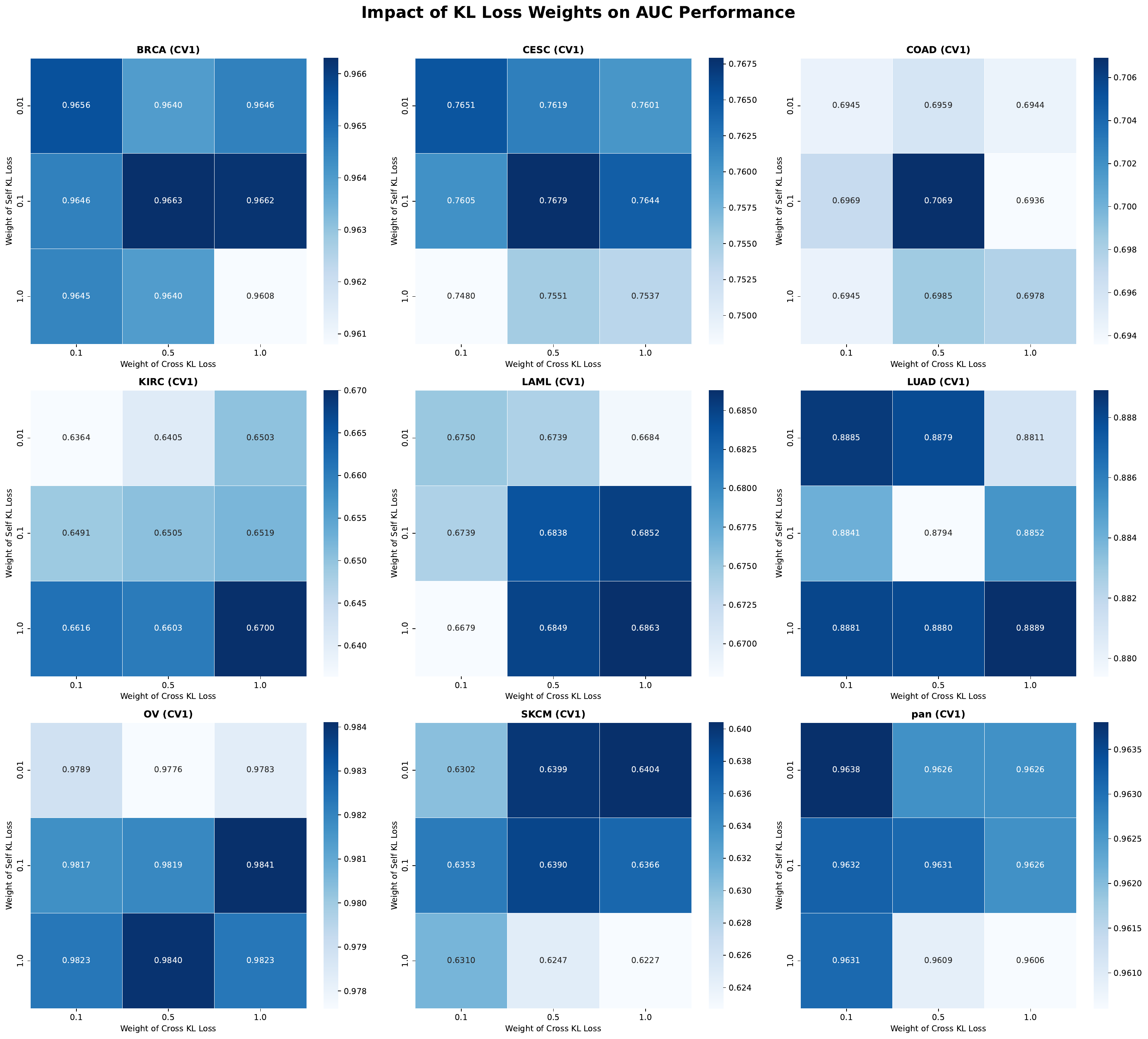}     \caption{
\textbf{Parameter sensitivity heatmap of the omics VAE module.} 
This shows the change in AUC on the 8 specific cancers and the pan-cancer datasets under the CV1 scenario 
for different combinations of $\lambda_{\text{self}}$ and $\lambda_{\text{cross}}$.
    }
    \label{fig:kl_heatmap}
\end{figure}

In the COAD dataset, although the default configuration of $\lambda_{\text{self}}=0.1, \lambda_{\text{cross}}=0.5$ 
achieved the highest performance (AUC=0.7069),
the model's performance remained stable within the narrow range of [0.693, 0.707] 
even when the parameters were increased or decreased by an order of magnitude, 
showing a fluctuation of less than 2\%.
In the OV dataset, this stability was even more significant, 
as the AUC for all parameter combinations remained at a high level between [0.977, 0.984].
This means that the model does not get stuck in a narrow local optimum, but instead learns robust feature representations.
The data trend shows that moderately increasing $\lambda_{\text{cross}}$ usually leads to a slight performance improvement.
This supports the design intuition of SynLeaF.
Giving a slightly higher weight to cross-modal alignment than to unimodal regularization 
helps force the model to discover common patterns among different omics data in the latent space.

In summary, SynLeaF is not sensitive to the hyperparameter settings of the VAE.
Therefore, we uniformly used the configuration of $\lambda_{\text{self}}=0.1$ and $\lambda_{\text{cross}}=0.5$ 
in all pan-cancer and single-cancer experiments.
This tuning-free characteristic greatly reduces the deployment cost when applying the model to new cancer types.

\subsection{The Best Topological Depth for RGCN}

In the knowledge graph encoder, the radius of the ``receptive field'' 
for a gene node within the biological network is governed by the quantity of RGCN layers $L$.
To determine the best topological range for capturing synthetic lethality interactions, 
we assessed the impact of different numbers of RGCN layers ($L \in \{1, 2, 3\}$) on the efficacy of the model across all datasets. 
Taking BRCA and LAML cancers as representatives for analysis, 
the trends in performance metrics illustrated in Supplementary Figure \ref{fig:gcn_layers} represent a general pattern that we observed.

\begin{figure}[!htbp]
    \centering
    \includegraphics[width=1\textwidth]{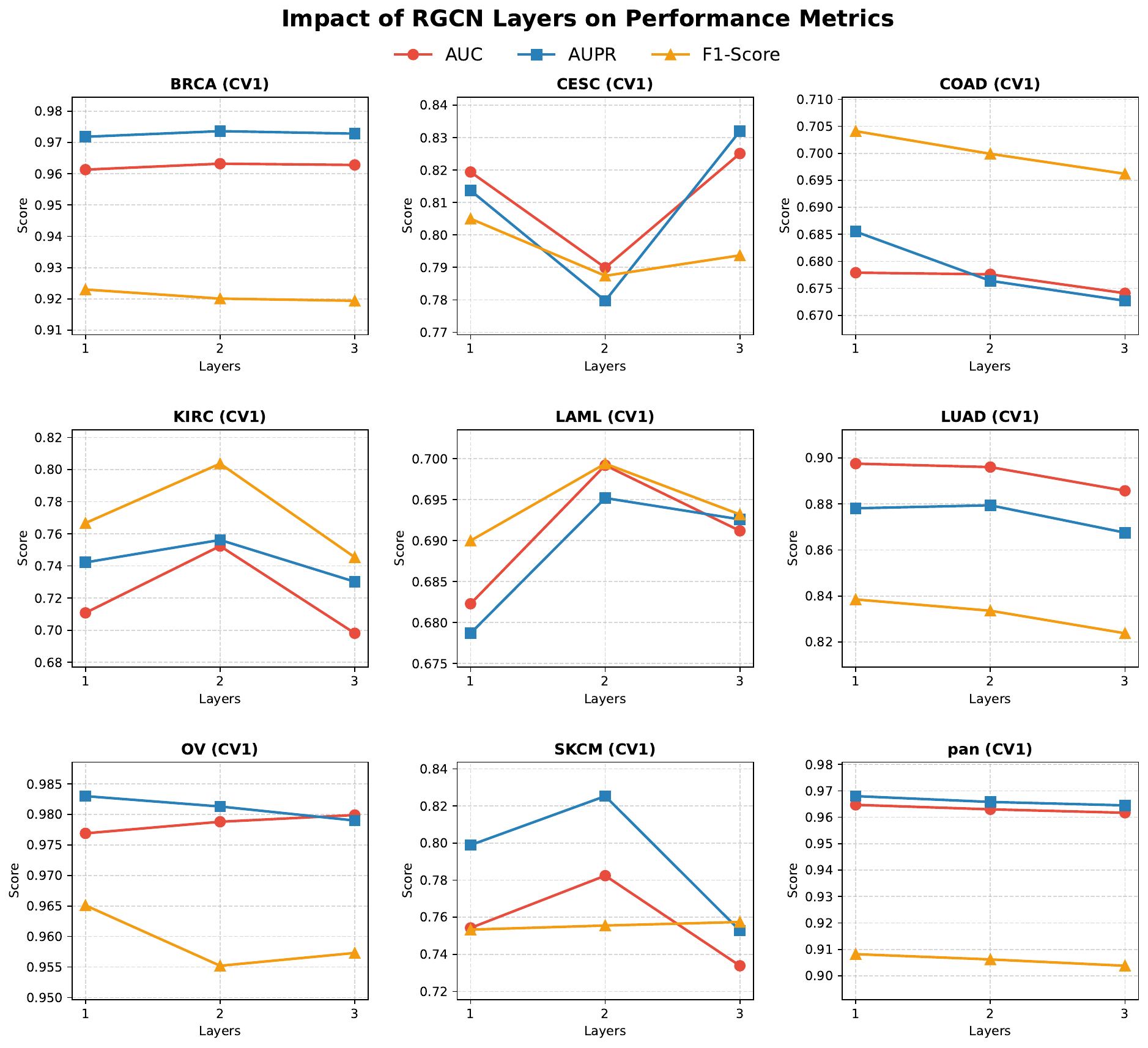} 
    \caption{
\textbf{Impact of RGCN layer count on model efficacy.} 
This figure illustrates the variations in Only KG model performance (AUC, AUPR, F1-Score) 
on the 8 specific cancers and the pan-cancer datasets across varying layer configurations.
    }
    \label{fig:gcn_layers}
\end{figure}

Empirical evidence indicates that a 2-layer RGCN structure usually achieves the best prediction performance.
Taking the LAML dataset as an example, 
the model AUC significantly increased from 0.6823 to 0.6992 
when the number of layers was increased from 1 to 2.
This indicates that only aggregating information from direct neighbors is often 
not enough to capture complex gene regulatory relationships.
Expanding the neighborhood range helps to introduce potential pathway-level contextual information.
The BRCA dataset also showed a similar trend, with the 2-layer structure achieving the best performance of 0.9632.

However, we consistently observed that when extending the depth to 3 layers,
the predictive capability on the majority of datasets declined (for example, LAML dropped back to 0.6912).
From a biological perspective, the synthetic lethality effect usually occurs 
within functionally closely related gene modules or local signaling pathways \cite{zinovyev2013synthetic}.
Furthermore, 2-hop neighbors usually already cover the range of the same pathway or directly interacting proteins.
On the other hand, neighbors that are 3 hops or more away 
are very likely to introduce functionally irrelevant noise nodes 
and dilute the key topological signal.
From a computational science perspective, deep graph neural networks can easily cause node representations to become homogenized, 
leading to the problem of over-smoothing \cite{nt2019revisiting}, 
thereby hindering the discrimination between features of distinct genes.
In addition, an increase in layer count results in an exponential expansion of the computation graph,
which leads to a significant increase in memory usage and training time.

In summary, choosing a 2-layer RGCN is the best balance 
between capturing long-range dependencies, avoiding noise interference, and controlling computational cost.
Therefore, SynLeaF uses a fixed number of graph convolutional layers of $L=2$ in all experimental settings.

\section{Experimental Design}

\subsubsection{Evaluation Metrics}

In this study, we employed three classic metrics for binary classification prediction: AUC, AUPR, and F1-score.

(1) \textbf{Area Under the ROC Curve} (AUC) quantifies the discriminative capacity of the model across the full range of thresholds,
identifying the relationship between the True Positive Rate (TPR) and the False Positive Rate (FPR).

(2) \textbf{Area Under the Precision–Recall Curve} (AUPR) maps Precision versus Recall 
and emphasizes performance in contexts characterized by data sparsity.
Particularly within highly imbalanced datasets, AUPR demonstrates the model's sensitivity in detecting positive instances more effectively than AUC.

(3) \textbf{F1-Score} constitutes the harmonic mean of precision and recall, 
serving as a comprehensive metric that balances both factors.
It achieves a high value exclusively when both precision and recall are substantial.

\subsubsection{Baseline Implementation}

In order to validate the efficacy of SynLeaF, 
we benchmarked our proposed framework against four existing state-of-the-art models for SL prediction, namely SLGNN, ELISL, PTGNN, and MPASL.
We referred to the corresponding papers and their official code repositories to reproduce these baseline models, and we adjusted the inputs to ensure that the datasets used were exactly consistent with this study, with detailed reproduction settings as follows.

(1) \textbf{SLGNN}: 
We adopted hyperparameter settings strictly consistent with the original paper: 
learning rate $lr=0.002$, similarity regularization weight $\lambda_1=1e-3$, and L2 regularization weight $\lambda_2=1e-4$.

(2) \textbf{ELISL}: 
The experiments used the default hyperparameters from the code without additional grid search, 
and the PPI data for ELISL was sourced from the STRING database \cite{szklarczyk2021string} with a version consistent with that provided in the code repository.

(3) \textbf{PTGNN}: 
Since the original authors did not provide the dataset, 
we referred to the method provided in the SL Benchmark \cite{feng2024benchmarking} to construct the required data.
The PPI data comes from BioGRID \cite{stark2006biogrid} (version 2022-06-25), and the GO data is from the version dated 2022-07-01.
Consistent with the SL Benchmark, we unified the input protein sequence length to 600, 
used the simUI method to calculate the semantic similarity of GO, 
and did not use random walk or the integration method of InteGO2 \cite{peng2016intego2}.

(4) \textbf{MPASL}: 
We adopted the default hyperparameter configuration from the code repository for the experiments.
To solve the problem where the model could not build a graph due to genes lacking known SL associations in CV2 and CV3 splits, 
we introduced a Void Node strategy to connect isolated genes to virtual nodes.
For the pan-cancer task, we used the default batch size of 512.
For single-cancer tasks, to mitigate NaN occurrences during training, the batch size was adjusted to 16. 
In the rare instances where NaN results persisted, they were treated as 0 during the computation of the mean and standard deviation.

\subsubsection{Hyperparameter Configuration}

To assess the model's capacity for generalization, 
we utilized fixed hyperparameter settings for both single-cancer and pan-cancer tasks
without performing grid search tuning for specific datasets.
Considering the differences in sample size between pan-cancer and single-cancer datasets, 
we set adapted parameter combinations for these two types of tasks, 
and the specific configurations are shown in Supplementary Table \ref{tab:hyperparams}.

\begin{table}[htbp]
    \centering
    \caption{Hyperparameter settings for Single-Cancer and Pan-Cancer tasks}
    \label{tab:hyperparams}
    \begin{tabular}{lcc}
        \toprule
        \textbf{Parameter} & \textbf{Single-Cancer} & \textbf{Pan-Cancer} \\
        \midrule
        Omics types & \text{cnv,exp,mut,myl} & \text{cnv,exp,mut} \\
        Omics VAE Hidden Dims & $[512, 256]$ & $[2048, 1024, 512, 256]$ \\

        Batch Size & 384 & 2048 \\
        Epochs & 200 & 30 \\
        \midrule
        Latent Dimension & \multicolumn{2}{c}{128} \\
        RGCN Layers & \multicolumn{2}{c}{2} \\
        RGCN Hidden Channels & \multicolumn{2}{c}{256} \\
        RGCN Output Channels & \multicolumn{2}{c}{128} \\
        Weight of Self KL Loss & \multicolumn{2}{c}{0.1} \\
        Weight of Cross KL Loss & \multicolumn{2}{c}{0.5} \\
        Weight of Distill Loss & \multicolumn{2}{c}{1} \\
        Learning Rate & \multicolumn{2}{c}{$1 \times 10^{-3}$} \\
        Weight Decay & \multicolumn{2}{c}{$1 \times 10^{-4}$} \\
        Dropout (VAE/Graph/MLP) & \multicolumn{2}{c}{0.2 / 0.5 / 0.5} \\
        \bottomrule
    \end{tabular}
\end{table}

For the omics encoder, the single-cancer task uses encoding layer dimensions of $[512, 256]$, 
while the pan-cancer task extends this to $[2048, 1024, 512, 256]$ to adapt to the increase in data volume.
For pan-cancer, since there is no methylation data in TCGA, there are only three omics types.
The latent variable dimensions for both are fixed at 128, and the dropout rate corresponding to the VAE layer is fixed at 0.2.
To balance the reconstruction error and the KL divergence regularization term, 
we set the KL divergence weight $\lambda_{\text{self}}$ to 0.1 and the weight $\lambda_{\text{cross}}$ to 0.5.

For the knowledge graph encoder, the RGCN is configured with 2 layers to capture local neighborhood information within a 2-hop range.
The node features of the RGCN are initialized with 128-dimensional random vectors, 
the hidden layer size is fixed at 256, while the output dimension is aligned with the omics encoder at 128.
To mitigate overfitting risks, we apply a dropout probability of 0.5 to the graph layer.
In the classifier part, the final MLP also utilizes a dropout rate of 0.5.

Regarding model optimization, we employed the RAdam optimizer \cite{liu2020variance} combined with the Lookahead mechanism \cite{zhang2019lookahead} using default parameters, 
configured with an initial learning rate of $1 \times 10^{-3}$ alongside a weight decay factor of $1 \times 10^{-4}$.
During the UMT stage of multimodal fusion, the coefficient $\lambda_{\text{distill}}$ for knowledge distillation was assigned a value of 1.
Regarding the training protocol, the maximum number of epochs was fixed at 200 for single-cancer tasks and 30 for the pan-cancer task.
During the training process, we retained the model checkpoint corresponding to the peak AUC score on the validation set,
utilizing it for the final performance assessment on the test set.

\subsubsection{Implementation Details}

All experiments in this study were completed on a high-performance computing server.
The hardware platform was equipped with a Nettrix X640 G40 server, 
featuring two Intel(R) Xeon(R) Platinum 8358 CPUs @ 2.60GHz (totaling 128 logical cores),
with 500GB of RAM and 7TB of hard disk storage.
Accelerated computing utilized 8 NVIDIA RTX A6000 graphics cards, each with a video memory capacity of up to 48GB, interconnected via NVLink for high speed.

Regarding the software environment, the model is implemented based on Python 3.11 and the PyTorch deep learning framework, 
while the graph neural network part utilizes the PyTorch Geometric (PyG) library \cite{fey2019fast}.
To improve training efficiency, we implemented Distributed Data Parallel (DDP) training using Hugging Face's Accelerate library \cite{wolf2020transformers}.
Each training task was automatically assigned to 2 graphics cards for parallel execution to fully utilize computational resources.
To eliminate the influence of randomness, we set fixed random seeds for all experiments, 
and the reported results are the averages of 5-fold cross-validation.

\section{Supplementary Tables}

\begin{table}[!htbp]
    \centering
        \caption{Statistics of synthetic lethality pair counts before data processing}
    \label{tab:sl_count_before}     \begin{tabular}{lccc}
        \toprule
Cancer Type & \# SL Genes & \# Pos. & \# Neg. \\
        \midrule
BRCA   & 1143	& 1444	 &   1037 \\
CESC   & 190	&  144	 &  4763 \\
COAD   & 18707	& 1728	 &  79323 \\
KIRC	&  92	&   60	 &   2514 \\
LAML   & 282	&  1191	 &  19308 \\
LUAD   & 914	&  596	 &  5516 \\
OV    & 89	&  253	 &  556 \\
SKCM   & 18719	& 107	 &  18702 \\
pan   & 9870	& 35815 &   3740 \\
        \bottomrule
    \end{tabular}
\footnotetext{\# Pos. denotes the count of positive samples; \# Neg. denotes the count of negative samples.}
\end{table}

\section{Supplementary Figures}

\begin{figure}[!htbp]
    \centering
    \includegraphics[width=1\textwidth]{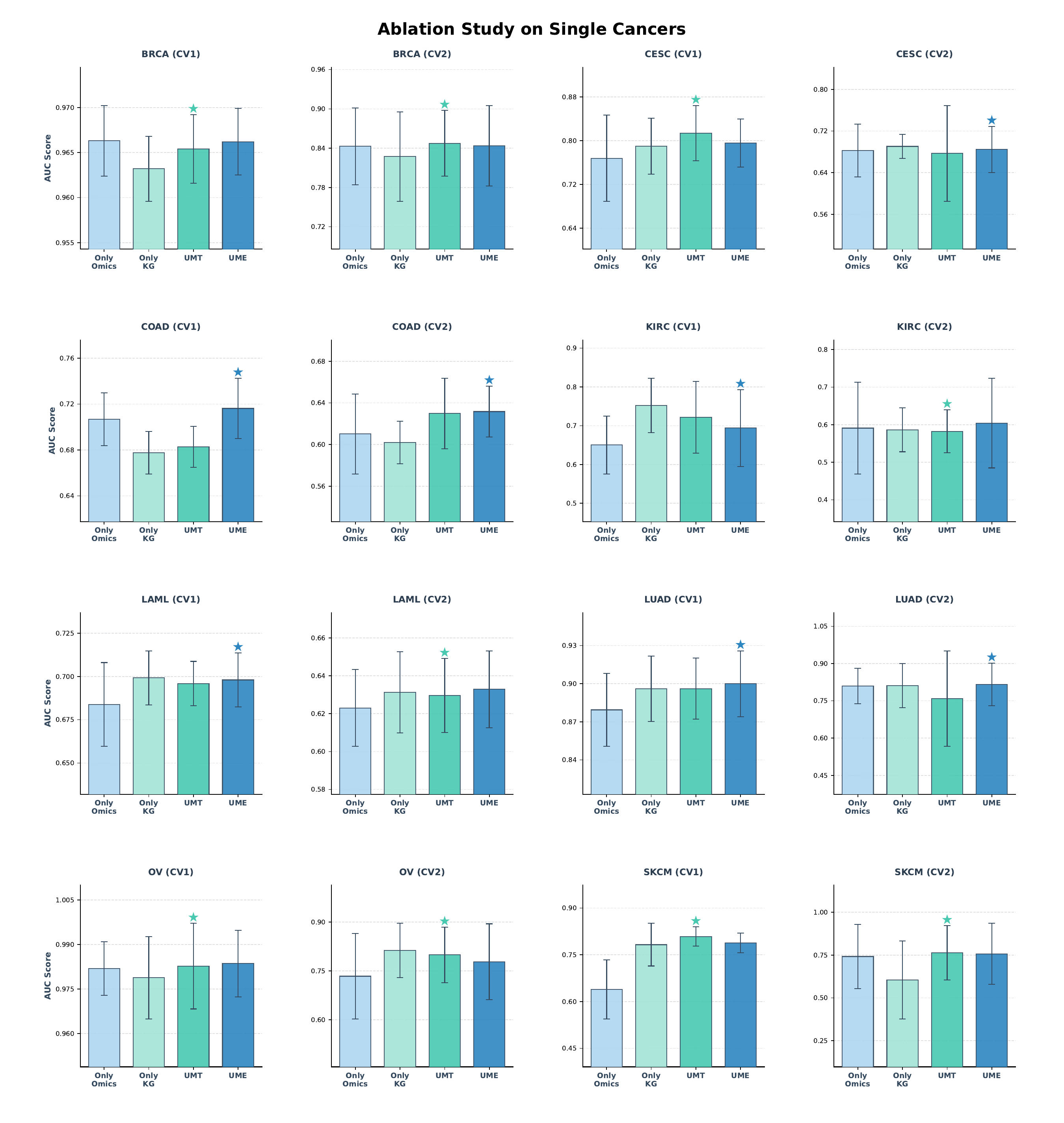}     \includegraphics[width=1\textwidth]{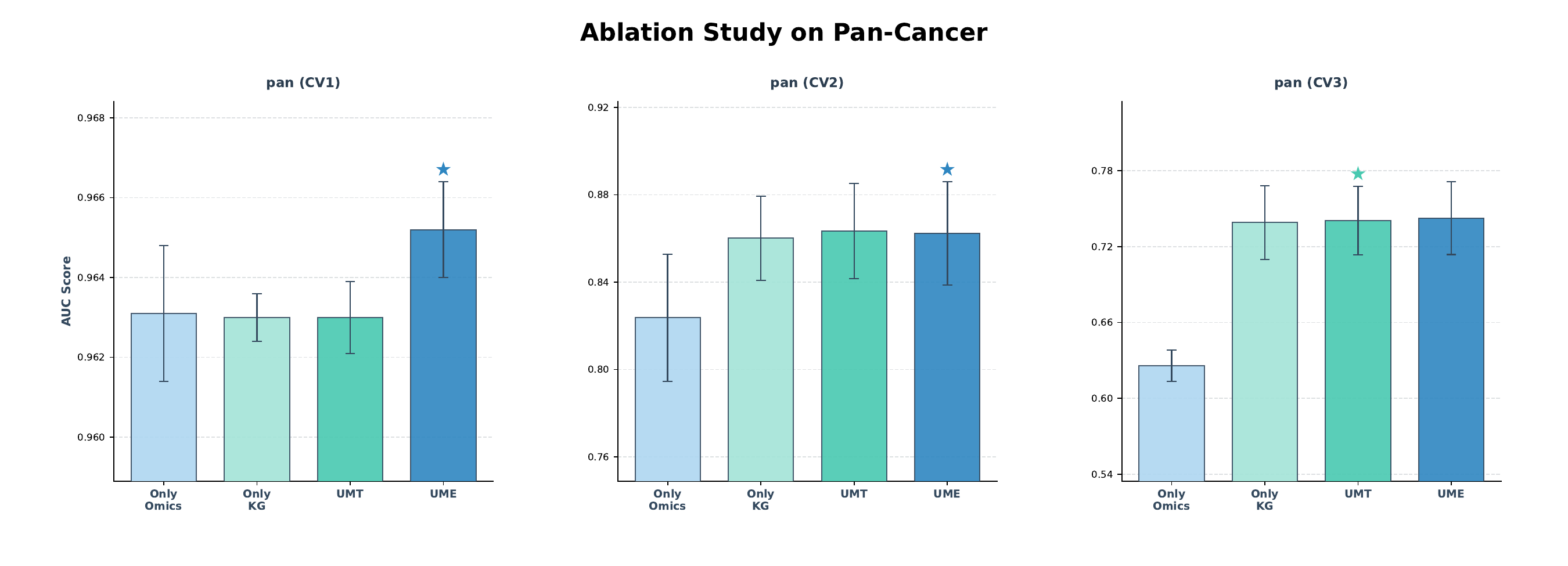}     \caption{
\textbf{Performance comparison of UMT and UME fusion strategies on different cancer datasets.}
This figure shows the AUC scores for the unimodal benchmark models (Only Omics, Only KG) 
and the two multimodal fusion strategies (UMT, UME)
across all single-cancer and pan-cancer datasets.
The height of the bars corresponds to the mean AUC obtained via 5-fold cross-validation, 
while the standard deviations are denoted by the error bars.
The star indicates the optimal fusion strategy 
that was finally selected by the SynLeaF adaptive selection mechanism
on that dataset.
        }
    \label{fig:ablation_bar} 
\end{figure}

\clearpage
\backmatter
